\documentstyle[eqsecnum,prd,aps,epsf]{revtex}

\begin{document}

\newcommand{\beq}{\begin{equation}}
\newcommand{\eeq}{\end{equation}}
\newcommand{\beqn}{\begin{eqnarray}}
\newcommand{\eeqn}{\end{eqnarray}}
\newcommand{\pa}{\partial}
\newcommand{\vp}{\varphi}
\def\zero{\hbox{$_{(0)}$}}
\def\bL{\hbox{$\,{\cal L}\!\!\!\!\!-$}}

\begin{center}
{\large\bf{Equilibrium and stability 
of supermassive stars in binary systems 
}}
~\\
~\\
Masaru Shibata$^{1}$, 
Stuart L. Shapiro$^{2,3}$, 
and 
K\=oji Ury\=u$^4$ \\
~\\
$^1${\em Graduate School of Arts and 
  Sciences,~University~of Tokyo,\\
Komaba, Meguro, Tokyo 153-8902, Japan}\\
$^2${\em Department of Physics, 
University of Illinois at Urbana-Champaign, Urbana, IL 61801, USA} \\
$^3${\em Department of Astronomy and NCSA, 
University of Illinois at Urbana-Champaign, Urbana, IL 61801, USA}
$^4${\em Department of Physics, University 
of Wisconsin at Milwaukee, Milwaukee, WI 53201, USA}
\end{center}

\begin{abstract}
We investigate the equilibrium and stability of supermassive 
stars of mass $M \agt 10^5M_{\odot}$ 
in binary systems. 
We find that corotating binaries 
are secularly unstable for close, circular orbits with 
$r \alt 4R(M/10^6M_{\odot})^{1/6}$ 
where $r$ is the orbital separation and $R$ the stellar radius. 
We also show that corotation 
cannot be achieved for distant orbits with 
$r \agt 12 R (M/10^6M_{\odot})^{-11/24}$, 
since the timescale for viscous angular momentum 
transfer associated with tidal torques is longer than the evolution timescale 
due to emission of thermal radiation. These facts suggest that the allowed 
mass range and orbital separation for 
corotating supermassive binary stars is severely restricted. 
In particular, for supermassive binary stars of 
large mass $M \agt 6\times 10^6M_{\odot}$, corotation 
cannot be achieved, as viscosity is not adequate to mediate the 
transfer between orbital and spin angular momentum. 
One possible outcome for binary supermassive stars is the onset of 
quasi-radial, relativistic instability which drives each star to collapse 
prior to merger: We discuss alternative outcomes of collapse and 
possible spin states of the resulting black holes. 
We estimate the frequency and amplitude of gravitational waves 
emitted during several inspiral and collapse scenarios. 
\end{abstract}


\section{Introduction}

Recent astronomical observations provide increasingly 
strong evidence 
that supermassive black holes of mass $\sim 10^6-10^{10}M_{\odot}$, 
exist and that they are the 
central engines of active galactic nuclei and quasars \cite{Rees}. 
In addition to their importance for a fundamental understanding of 
active galactic nuclei and quasars, gravitational waves from 
the formation of the supermassive 
black holes and from inspiraling supermassive binary black holes 
are likely sources for proposed laser interferometric 
gravitational wave detectors in space, such as the Laser 
Interferometer Space Antenna (LISA) \cite{KIP0,LISA}. 
However, the scenario by which supermassive black holes form is 
still uncertain. Viable stellar dynamical and hydrodynamical routes 
leading to the formation of supermassive black holes have been proposed 
(see, e.g., \cite{BR}). In one hydrodynamical scenario, a supermassive 
gas cloud is build up from the multiple collisions of stars or 
small gas clouds 
in stellar clusters to form a supermassive star. Supermassive stars 
subsequently collapse to black holes after 
quasi-stationary contraction as a result of cooling \cite{ZN,ST}. 
Such a scenario for forming a supermassive star and black hole 
has been investigated in detail 
in the 1960s and 70s using simple analytical models or assuming 
spherical symmetry (see, e.g., \cite{BS} for a review and 
references). 

Supermassive stars are likely to be rapidly rotating (see, e.g., \cite{LR}). 
For studies of rapidly rotating stars in equilibrium, 
it is necessary to perform numerical computations using modern 
numerical techniques \cite{ROTSTAR} which were not feasible until 
quite recently. 
Baumgarte and Shapiro recently performed a detailed numerical analysis 
of a rapidly rotating supermassive star \cite{BS}. 
Because of the fact that the viscous or magnetic 
braking timescale for angular momentum 
transfer is likely to be shorter than the evolution timescale of supermassive 
stars in the typical mass range, the star can be assumed to be in 
rigid rotation at the mass-shedding limit 
(which implies that the spin angular velocity of each star is 
equal to the Kepler velocity) \cite{KS}. They find that at 
the onset of quasi-radial collapse, the non-dimensional spin parameter 
$cS/GM^2$, where $S, M, c$ and $G$ are 
spin, mass, light velocity, and gravitational constant, is 
nearly equal to unity, so that rotating supermassive stars 
might not collapse directly to a black hole. 
They speculate that the collapse may be inhibited, forming a disk 
and/or a bar, or possibly fragmenting into several blobs. 
If such non-axisymmetric structures form during the collapse, 
quasi-periodic gravitational waves of $\sim 10^{-4}-10^{-2}$Hz may be 
emitted and provide a strong source for laser interferometeric 
gravitational wave detectors in space \cite{KIP0}.

In this paper, we consider supermassive stars in binary systems. 
Such binaries may arise naturally from fragmentation of a large 
gas cloud with rapid rotation, or during multiple mergers 
of small gas clouds and/or stars \cite{BR,LR}. 
Even in binary systems, each supermassive star is likely to reach 
quasi-radial instability eventually after quasi-stationary contraction. 
In binaries, most of the angular momentum of the system can be distributed 
in orbital motion so that the spin of each star may be much smaller 
than that for isolated supermassive stars. 
In particular, small spin can be achieved for the case 
when viscous transport 
between orbital and spin angular momentum is effective. 
If so, each star is in slow rotation with respect 
to the inertial frame, and collapse to a 
black hole may proceed in an almost spherical manner, with tiny 
emission of gravitational waves. 
Thus, the scenario for forming 
black holes in binary systems could be considerably 
different from that for isolated supermassive stars. 
Also, a binary containing supermassive stars can be a progenitor 
of an inspiraling, supermassive black hole binary in a close orbit.
Such a binary can emit gravitational waves in the frequency range 
appropriate for LISA [see Eq. (\ref{frequency})]. 
To detect gravitational waves from binaries using matched filtering 
technique \cite{KIP,CF}, it is useful to anticipate possible mass 
and spin states of the black holes. 
With this motivation, we investigate the equilibrium and 
stability of supermassive binary stars in this paper.

The paper is organized as follows.
In Sec. II, we estimate various timescales which are relevant for 
evolution of supermassive binary stars.
In Sec. III, we construct analytical equilibrium models 
of corotating, supermassive binary stars and investigate their 
stability. In Sec. IV, we present numerical solutions for 
corotating, supermassive binary stars to demonstrate that 
the analytical results shown in Sec. III are qualitatively correct. 
In Sec. V, we consider the evolution of corotating, 
supermassive binary stars as a result of thermal radiation, and clarify 
their final fate. In Sec. VI, we 
briefly discuss gravitational waves from supermassive stars 
in the present scenarios. Section VII is devoted to a brief summary. 
Hereafter, we adopt the geometrical unit in which $G=c=1$.

\section{Timescales}

In this paper, we investigate equilibrium and stability of supermassive 
stars in close binary systems. Before proceeding, we estimate 
the relevant timescales characterizing such systems 
to clarify possible states of supermassive binaries. 
Here, we define supermassive stars to be self-gravitating objects of mass 
larger than $\sim 10^5M_{\odot}$ (see discussion below).
Also, we assume that the binary 
is an isolated system; namely, the accretion of mass and 
heating from outside are assumed to be negligible. 

The equation of state for supermassive stars is dominated 
by thermal radiation pressure with a small correction due to 
baryonic gas pressure. Such stars are 
convective with constant entropy per baryon \cite{ZN,ST,LR}. 
These conditions imply that the structure of the 
supermassive stars is very close to that of an $n=3$ polytrope
so that 
\beq
P=K\rho^{4/3} + P_g,~~~P_g \ll K \rho^{4/3}, 
\eeq
where $P$, $\rho$, $P_g$, and $K$ are the total 
pressure, baryon mass density, gas pressure, and 
polytropic constant, respectively; $K$ is approximately related to the 
entropy of radiation, $s_r$, according to \cite{ZN,ST} 
\beq
K={a \over 3}\biggl({3s_r \over 4m_{\rm H} a}\biggr)^{4/3}. 
\eeq
Here, $a$ is the radiation density constant, 
and $m_{\rm H}$ the mass of hydrogen; 
we assume that supermassive stars are composed of 
pure hydrogen for simplicity. 
For $n=3$ spherical polytropes in Newtonian theory, 
the mass $M$ is uniquely determined by $K$ according to 
\beq
M=\biggl({k_1 K \over k_2} \biggr)^{3/2},
\eeq
where $k_1$ and $k_2$ are constants of order unity associated with the 
Lane-Emden function (see Sec. III). Using this relation, 
$s_r$ can be written approximately as \cite{ST}
\beq
{s_r \over k} \simeq 0.942\biggl({M \over M_{\odot}}\biggr)^{1/2},
\eeq
where $k$ denotes the Boltzmann constant. 

In Newtonian theory, pure 
$n=3$ spherical polytropes are marginally stable to radial 
collapse, and with a tiny general relativistic (GR) correction, 
they are unconditionally unstable. 
To stabilize such a star, some additional perturbations are necessary. 
One perturbation is spin angular momentum. As shown in \cite{BS}, 
supermassive stars are stabilized if they have 
a large spin angular velocity. However, 
in this paper, we mainly analyze corotating, supermassive 
binary stars, whereby the spin angular momentum of each star 
is not large enough to stabilize the star against radial collapse. 
To stabilize such slowly rotating supermassive stars, the role of 
baryonic gas pressure, which constitutes a small fraction of 
the total pressure, is crucial. 

The evolution of isolated supermassive stars that do not 
have large angular momentum and are stabilized by 
the gas pressure has been 
investigated in detail many years ago \cite{ZN,ST}. According to 
a typical scenario, a supermassive star shrinks quasi-statically 
as a result of emission of the thermal radiation 
from its surface until it reaches a critical point where 
a GR instability sets in against radial collapse. 
The radial instability occurs 
for a critical compactness \cite {chandra} [see also Eq. (\ref{compactness})] 
\beq
{M \over R} > 0.445\biggl(\Gamma-{4 \over 3}\biggr),
\eeq 
where $R$ is the circumferential radius, and 
$\Gamma$ is the effective adiabatic constant. Here, $\Gamma$ may be 
approximately written as 
\beq
\Gamma \simeq {4 \over 3}+{\beta \over 6}, 
\eeq
where $\beta$ is 
the ratio of the gas pressure to the 
radiation pressure, $\beta\equiv P_g/(K\rho^{4/3}) \ll 1$, which 
can be expressed as \cite{ST}
\beq
\beta = {8 k \over s_r} \simeq 
8.5 \biggl({M \over M_{\odot}}\biggr)^{-1/2}. \label{eqbeta}
\eeq
Thus, the constraint of $M/R$ for 
a stable supermassive star of mass $M$ can be rewritten as 
\beq
{M \over R} \alt 0.63 \biggl({M \over M_{\odot}}\biggr)^{-1/2}.
\label{compactconstr}
\eeq

In this evolutionary scenario, there are two relevant timescales; the 
emission timescale of thermal radiation and the 
dynamical timescale. Supermassive stars emit the thermal radiation 
from the stellar surface with the Eddington luminosity 
\beq
L_{\rm Edd}={4\pi M \over \kappa_{\rm T}},
\eeq
where $\kappa_T$ is the opacity with respect to the Thomson scattering. 
The total energy of a supermassive star is approximately \cite{ZN,ST}
\beq
E \simeq -\beta {M^2 \over R}. 
\eeq
Thus, the emission timescale of thermal radiation is 
\beq
\tau_{\rm evol}\equiv
{|E| \over L_{\rm Edd}} \sim 1.4\times 10^{10} {\rm sec} 
\biggl({M/R \over 10^{-3}}\biggr)
\biggl({\beta \over 10^{-3}}\biggr),
\eeq
where, in evaluating $\kappa_{\rm T}$, 
we assume pure ionized hydrogen. 
The dynamical timescale is approximately 
\beq
\tau_{\rm dyn} \equiv \rho_c^{-1/2}\simeq 1.4 \times  
10^4 {\rm sec} \biggl({M/R \over  10^{-3}}\biggr)^{-3/2} 
\biggl({M \over  10^6M_{\odot}}\biggr), 
\eeq
where $\rho_c$ denotes the maximum (central) density of supermassive stars 
which, for $n=3$ spherical polytropes, 
can be expressed as $\rho_c \simeq 12.94M/R^3$ 
(see next section and Table I).

For a star to evolve quasi-statically, 
the inequality, $\tau_{\rm evol} > \tau_{\rm dyn}$, 
must be satisfied, so that 
\beq
{M \over R} > 1.7\times 10^{-6} \biggl({M \over 10^6M_{\odot}}\biggr)^{3/5},
\eeq
where we use Eq. (\ref{eqbeta}) for evaluating $\beta$. 
From the requirement of stability, $M/R$ also has to satisfy 
Eq. (\ref{compactconstr}). Thus, for existence of a stable 
supermassive star in quasi-static evolution, the mass has to be less than 
$\sim 2\times 10^8M_{\odot}$. 
If the mass is larger than this value, supermassive stars 
collapse dynamically without evolving through a quasi-static phase. 
Furthermore,  supermassive stars of mass 
$M \alt 10^5 M_{\odot}$ reach sufficiently high temperature 
for nuclear burning to become important before reaching 
the onset of the instability. 
If nuclear reactions set in, the temperature 
increases so rapidly that the star undergoes a violent 
explosion \cite{Fricke}. Thus, 
we restrict our attention to 
stars with $M \agt 10^5 M_{\odot}$ which are stable until 
they reach the onset of radial instability.

For binary systems, several 
additional timescales have to be taken into account. One is 
the radiation-reaction timescale of gravitational waves, which is 
written using the quadrupole formula as \cite{ST2}
\beq
\tau_{\rm GW} \simeq 
{5 \over 512}{r^4 \over M^3} \simeq 2\times 10^{14} {\rm sec}
\biggl( {r/M \over 10^3} \biggr)^4 \biggl({M \over 10^6 M_{\odot}}\biggr),
\label{taugw}
\eeq
where we assume that the mass of each star, $M$, is identical 
(i.e., the total mass is $2M$), and where 
$r$ denotes the orbital separation between two centers of mass. 
Since $r/M > R/M \agt 10^3$ for 
stable supermassive stars with $M \geq 10^5M_{\odot}$, 
we immediately find $\tau_{\rm GW} \gg \tau_{\rm dyn}$ and 
$\tau_{\rm GW} \gg \tau_{\rm evol}$, so 
that gravitational radiation reaction is not relevant in this problem. 

According to standard scenarios, 
supermassive stars were formed near galaxy centers 
in the early stage of galaxy formation \cite{BR}. 
Associated with galaxy formation in its early stage, 
gas may settle toward the center of the galaxies. 
If the gas accretes onto supermassive binary stars, 
it can be a source of friction which can 
decelerate the orbital motion of the stars. 
The dragging timescale can be written crudely as
\beq
\tau_{\rm drag} \sim {M \over \rho_0 R^2 r \Omega} 
\sim  {4\pi \over 3}\biggl({\rho_c \over \rho_m}\biggr)^{1/2}
\biggl({\rho_m \over \rho_0}\biggr)
\biggl({r \over R}\biggr)^{1/2}
\tau_{\rm dyn},
\eeq
where 
$\rho_0$ and $\rho_m$ denote the typical density of the accreting gas, 
and the mean density of supermassive stars. 
If $\tau_{\rm drag} < \tau_{\rm evol}$ 
the orbital decay due to 
friction can become effective in inducing merging. 
However, to achieve this condition, $\rho_0$ has to be extraordinary 
large, 
\beq
\rho_0 \agt 10^{-9} {\rm g/cm}^3
\biggl({M/R \over 10^{-3}}\biggr)^{1/2}
\biggl({M \over 10^{6}M_{\odot}}\biggr)^{-1/2}
\biggl({r \over 2R}\biggr)^{1/2},
\eeq
implying that $\rho_0$ has to be larger than 
$\sim 10^{13}M_{\odot}/{\rm pc}^3$. 
Also, the total amount of mass of the accreted gas 
in time $\tau_{\rm evol}$ has to be as large as $\sim M$ so that 
the binary in such situation cannot be regarded as 
an isolated system contrary to 
our assumption in Sec. II. 
Thus, we do not consider the effect of the friction.

The viscous timescale plays an important role in determining 
the velocity field in binary systems. 
Here, we have to take into account two timescales. 
One is associated with the tidal torque from the companion star 
(hereafter,we refer to this viscous timescale as $\tau_{\rm vis}$). 
If $\tau_{\rm vis}$ is shorter than $\tau_{\rm evol}$,  
transfer between orbital and spin angular momentum can be 
efficient and corotation is achieved. 
On the other hand, for $\tau_{\rm vis} > \tau_{\rm evol}$, 
angular momentum transfer is not important, 
implying that the spin of each star is 
determined solely by the viscous angular momentum transfer inside each star 
(hereafter, we refer to this timescale as $\tau_{\rm vis0}$). 
In the case when $\tau_{\rm vis} > \tau_{\rm evol} > \tau_{\rm vis0}$, 
each star will be in rigid rotation in its comoving 
frame. For $\tau_{\rm vis} > \tau_{\rm vis0} > \tau_{\rm evol}$, 
vorticity of stars is conserved, but such a high value of 
$\tau_{\rm vis0}$ is unlikely [see below Eq. (\ref{eq223})].

The timescale for the viscous transfer of the angular momentum induced 
by the tidal torque is approximately \cite{tassoul}
\beq
\tau_{\rm vis} \sim {R^2 \over \nu}\biggl({r \over R}\biggr)^6,
\eeq
where $\nu$ is the viscous parameter. 
As often found in astrophysical contexts, 
microscopic, molecular viscosity yields timescales that are larger than 
the evolutionary timescale by many orders of magnitude. However, 
the effect of turbulent viscosity can be estimated 
by assuming that the velocity of turbulent motion $v_t$ is 
an appreciable fraction of the velocity of sound \cite{SS},
\beq
v_t = \alpha v_{\rm sound},
\eeq
where we take the dimensionless viscosity parameter $\alpha$ 
to be in the range $0.01 \alt \alpha \alt 1$. 
Assuming the characteristic length scale of turbulence to be 
of $O(R)$, we can estimate $\nu$ as
\beq
\nu \sim \alpha R v_{\rm sound}. 
\eeq
Then, 
\beq
\tau_{\rm vis} \sim {R \over \alpha v_{\rm sound}}
\biggl({r \over R}\biggr)^6
\sim \alpha^{-1} \biggl({r \over R}\biggr)^6 \tau_{\rm dyn}.
\eeq
To achieve corotation, 
the condition $\tau_{\rm vis} < \tau_{\rm evol}$ has to apply, 
and hence, 
\beq
{r \over R} \alt 14 \alpha^{1/6}\biggl({M/R \over 10^{-3}}\biggr)^{5/12}
\biggl({M \over 10^{6}M_{\odot}}\biggr)^{-1/4}. \label{constrr}
\eeq
Since $M/R$ has to satisfy Eq. (\ref{compactconstr}), we obtain 
\beq
{r \over R} \alt 12 \alpha^{1/6}
\biggl({M \over 10^{6}M_{\odot}}\biggr)^{-11/24}. \label{constr0}
\eeq
The existence of binary systems requires $r$ to be larger than $\sim 2R$, 
implying 
\beq
M < 5\times 10^7 \alpha^{4/11} M_{\odot} . \label{constr1}
\eeq 
Therefore, supermassive stars in corotating binaries can exist 
only for close orbits [cf. Eq. (\ref{constrr})], 
for compact stars [cf. Eq. (\ref{constr0})], 
and for a fairly small mass [cf. Eq.(\ref{constr1})]. In particular, 
supermassive stars 
in binary systems with masses larger than $\sim 10^8M_{\odot}$ 
cannot be in corotation. In the next section, we will 
show that the stability may further restrict 
the allowed region for these parameters.

Even if tidal interactions are not important, the viscous 
transfer of angular momentum 
inside each star can be important 
\cite{BS}, since the viscous timescale is 
\beq
\tau_{\rm vis0} \sim \alpha^{-1} \tau_{\rm dyn}.\label{eq223}
\eeq
Thus, as long as 
$M/R > 1.7\times 10^{-6} \alpha^{-2/5}(M/10^6M_{\odot})^{3/5}$, 
the relation $\tau_{\rm vis0} < \tau_{\rm evol}$ always holds.

\section{Analytical model for corotating binary}

We investigate the stability of supermassive binary stars 
supported by a polytropic equation of state
\beq
P=K\rho^{\Gamma}.\label{pressure}
\eeq
In this section, $\Gamma$ is chosen to be close to $4/3$ according to  
\beq
\Gamma={4 \over 3} + \epsilon,~~~~~~~~~~\epsilon={\beta \over 6} \ll 1. 
\eeq
With this choice of equation of state, we can 
investigate the quantitative properties 
of supermassive stars of mass $M \gg M_{\odot}$ analytically by means of 
a variational principle. As mentioned in Sec. II, 
such stars have an effective adiabatic constant given approximately by 
\beq
\Gamma \simeq 
{4 \over 3} + 1.42\biggl({M \over M_{\odot}}\biggr)^{-1/2}.\label{eq142}
\eeq
If we assume that the mass of supermassive stars 
is in the range between $10^5 M_{\odot}$ and 
$2 \times 10^8 M_{\odot}$, $\epsilon$ should be in the range 
between $\sim 10^{-4}$ and $\sim 5\times 10^{-3}$ 
[Eq. (\ref{eqbeta})]. 

We here consider corotating supermassive stars of equal mass for 
simplicity. The alternative 
extreme case of binary component with large mass ratios is 
also investigated in Appendix A using a Roche binary model. There we  
show that the results are qualitatively the same.  
To determine equilibrium and stability of supermassive stars 
in corotating binary systems, we write the total energy for each star 
as the sum of the internal energy $U$, the potential energy $W$, 
the orbital kinetic energy $T$, the spin kinetic energy $T_s$ and 
the binding energy between two stars $W_b$ as \cite{ST,LRS} 
\beqn
&& U=k_1 K M x^{n/3},\\
&& W=-k_2 M^{5/3}x-k_4 M^{7/3}x^2,\\
&& T={M \over  8} r^2 \Omega^2,\\
&& T_s={1 \over 2} I\Omega^2,\\
&& W_b=-{M^2 \over 2 r}. 
\eeqn
Here, $M$ is the mass of each star, 
$x=\rho_c^{1/3}$ the density parameter 
($\rho_c$ is the maximum density again), 
$r$ the orbital separation between the 
two centers of mass, $n [=1/(\Gamma-1)]$ 
the polytropic index, $\Omega$ the orbital angular velocity, 
and $I$ the moment of inertia. We do not distinguish 
the rest mass (baryon mass) 
and gravitational mass since their difference is quite small 
and unimportant for supermassive stars. 
The nondimensional structure constants 
$k_1$, $k_2$, and $k_4$ are constants dependent on $n$ 
calculated from the Lane-Emden function \cite{ST}. 
In Table I, we list these constants for several $n$ near 3. 

In $W$, we include the first post-Newtonian correction 
taking into account GR effects on the stability against 
quasi-radial collapse. 
We neglect the first post-Newtonian corrections in $W_b$ since their 
magnitude is smaller than that of the Newtonian terms 
by several orders of magnitude as $M/r < M/2R \ll 1$.  
We also neglect the tidal energy in $W_b$ and $W$ \cite{LAI}. 
As we estimate the magnitude of this effect 
in Appendix B, it is not substantial for corotating binaries. 
As long as $\epsilon \agt 7\times 10^{-4}$, it is not important 
even in non-spinning binaries. 

The quantity $I$ may be written in the form 
\beq
I=\kappa M^{5/3}\rho_c^{-2/3},
\eeq
where $\kappa$ is a constant calculated from the 
Lane-Emden function (cf. Table I). 

The total energy per one star, $E$, then becomes 
\beq
E=k_1KM x^{3/n} - k_2M^{5/3}x - k_4M^{7/3}x^2 
+\biggl({Mr^2 \over 8}+{I \over 2}\biggr)\Omega^2 -{M^2 \over 2r}. 
\eeq
The equilibrium angular velocity 
$\Omega$ is derived from the relation $\pa E/\pa r|_{M,J,K}=0$ as 
\beq
\Omega=\sqrt{{2M\over r^3}},
\eeq
where $J$ is the angular momentum of one star 
\beq
J={Mr^2\Omega \over 4}+I\Omega
=(\eta\Omega^{-1/3}+\kappa x^{-2}\Omega)M^{5/3},\label{eqJ}
\eeq
and $\eta=2^{-4/3}$. 
We note that the second term denotes the spin of each star $S (=I\Omega)$. 

Substituting $r=(2M/\Omega^2)^{1/3}$ in $E$ and 
taking the first derivative of $E(=E[x,\Omega(x)])$ with respect to $x$, 
fixing $M$, $J$ and $K$, yields a second condition for equilibrium, 
\beqn
0={\pa E \over \pa x}\Big|_{M,J,K}={3k_1 \over n}K M x^{3\epsilon} 
- k_2 M^{5/3}-2k_4 M^{7/3}x 
+{\kappa \Omega^2 M^{5/3} \over x^3},\label{dedx}
\eeqn
where we use $\epsilon=1/n-1/3$ and the relation 
\beq
{\pa \Omega \over \pa x}\Big|_{M,J}={2\kappa \over x^3}\Omega\biggl(
-{\eta \over 3} \Omega^{-4/3}+{\kappa \over x^2}\biggr)^{-1} < 0.
\label{dwdx}
\eeq
The second derivative of $E(=E[x,\Omega(x)])$ with respect to $x$ becomes 
\beqn
{\pa^2 E \over \pa x^2}\Big|_{M,J,K}
=&&{9\epsilon k_1 \over n}K M x^{3\epsilon - 1}
-2k_4M^{7/3}
-{3\kappa M^{5/3}\Omega^2 \over x^{4}}
\biggl({\eta x^2 + \kappa\Omega^{4/3} \over \eta x^2 -3\kappa\Omega^{4/3}}
\biggr) \nonumber \\
=&&
3\epsilon k_2 M^{5/3}x^{-1} -2(1-3\epsilon)k_4M^{7/3}
-{3\kappa \Omega^2 M^{5/3} \over x^{4}} \biggl(
{\eta x^2 + \kappa \Omega^{4/3} \over \eta x^2 - 3\kappa \Omega^{4/3}}
+\epsilon \biggr), \label{d2Ed2x}
\eeqn
where to derive the second line, we use Eq. (\ref{dedx}).
For stable equilibrium, this second derivative has to be 
positive, and the zero point marks the onset of instability \cite{ZN,ST}. 

It is convenient to analyze equilibrium and stability 
fixing $\Omega^2/\rho_c=\Omega^2/x^3 \equiv \omega^2$, which is 
equivalent to fixing the ratio of the orbital separation to the 
stellar radius of stars: 
\beqn
\omega^2={8\pi \over  3}\biggl({R \over r}\biggr)^3
\biggl({\rho_c \over \rho_m}\biggr)^{-1}
\sim 0.02\biggl({2R \over r}\biggr)^3, \label{limit}
\eeqn
where $\rho_m$ denotes the mean stellar density, i.e., $3M/(4\pi R^3)$, and 
$\rho_c/\rho_m \sim 54$ for $\epsilon \ll 1$ (see Table I). 
Since $r > 2R$ for binary systems, 
$\omega^2$ should be less than $\sim 0.02$. 
(In reality, $\omega^2$ is less than $\sim 0.01$ since the 
equator of each star is significantly 
enlarged by the centrifugal force 
associated with the spin for close binaries.) 

The equilibrium condition 
$\pa E/\pa x|_{M,J,K}=0$ yields the relation between $M$ and $x$ as 
\beq
M=\biggl[{-(k_2-\kappa\omega^2)+\sqrt{(k_2-\kappa\omega^2)^2+24 k_1 k_4 K 
x^{1+3\epsilon}/n} \over 4k_4 x}\biggr]^{3/2}.\label{mass}
\eeq 
The condition for the onset of instability 
$\pa^2 E/\pa x^2|_{M,J,K}=0$ with Eq. (\ref{mass}) gives the 
solution of $x$ at the critical points as 
\beq
x =\biggl[
{3(\epsilon k_2-\kappa\omega^2 F_0)\{k_2-\kappa\omega^2 (1+3F_0-3\epsilon)\}
\over 2(1-3\epsilon)^2 (1+3\epsilon)k_1 k_4 K} \biggr]^{n/3}
\equiv x_c, \label{answerx}
\eeq
where 
\beq
F_0={\eta + \kappa \omega^{4/3} \over \eta -3 \kappa \omega^{4/3}}+\epsilon.
\eeq

Since our purpose here is to determine the stability, 
we restrict our attention to the case where $K$ is constant.
Equation (\ref{answerx}) implies that 
on every equilibrium curve, $M(x)$, of a fixed value of $\omega$, 
there exists one turning point $x_c$ for a sequence along which 
$J$ is constant as 
\beq
J=(\eta\omega^{-1/3}+\kappa \omega)x_c^{-1/2}M(x_c)^{5/3}. 
\eeq
However, the implications of the existence of 
turning points for small $\omega$ and large $\omega$ 
are different. 
Consider the relation between $\bar M = MK^{-n/2}$ and 
$(\bar \rho_c)^{(3-n)/2n}~(\bar \rho_c=\rho_c K^n)$ 
for $\epsilon=0.003$ in Fig. 1 as an example. 
We remark that the qualitative behavior is independent of $\epsilon$ 
as long as $0< \epsilon \alt 0.01$. 
In the presence of a general relativistic effect or 
in the case $\epsilon > 0$, $MK^{-n/2}$ is not constant even in 
the infinite separation (i.e., $\omega^2=0$), but 
it is still close to $\sim 4.5$ for a wide range of the 
central density and orbital separation as long as $\epsilon \ll 1$. 
The solid curves denote the relation between $\bar M$ and $\bar \rho_c$ 
for $\omega^2=0$, 0.005 and 0.01, and the dotted curve  
for the sequence of turning points defined by Eqs. (\ref{mass}) and 
(\ref{answerx}). We also plot the curves 
of constant $\bar J$ where $\bar J=J K^{-n}$ (the dashed lines). According 
to the turning point theorem \cite{TP}, 
a change of the sign of $d\bar M/d\bar\rho_c$ 
along a curve of constant value of $\bar J$ 
indicates the change of (secular) stability. 
Along curves of constant $\bar J$, there are in general 
two turning points for different value of $\omega$ 
at $\bar\rho_c=\bar\rho_1$ and $\bar\rho_2 (> \bar\rho_1)$ 
for $\bar J \agt 420$. The turning point at $\bar\rho_c=\bar\rho_1$ exists 
only for $\omega \geq \omega_{\rm crit}$ and 
that of $\bar\rho_2$ for $\omega \leq \omega_{\rm crit}$. 
Here, numerical calculation yields the approximate relation 
\beq
\omega_{\rm crit}^2 \simeq (0.35-0.38) \epsilon,~~~~
{\rm for}~~10^{-4} \leq \epsilon \leq 5 \times 10^{-3} .\label{newcon}
\eeq 
We can determine the stability for stars 
along a curve of constant $\bar J$ in the following manner. 
All the curves of constant $\bar J$ 
in Fig. 1 asymptotically approach the unstable branch of isolated 
spherical stars (the solid curve with $\omega^2=0$). 
Stars on this branch are unstable against radial collapse. 
This implies that a binary 
star with $\bar\rho_c > \bar\rho_2$ is secularly unstable against 
quasi-radial collapse. Since the stability changes at each 
turning point, a star with $\bar\rho_1 < \bar\rho_c < \bar\rho_2$
is stable and one with $\bar\rho_c < \bar\rho_1$ is secularly unstable. 

The compactness of star at $x=x_c$ is evaluated as 
\beq
{M \over R}={3\epsilon k_2 \over 2 (1-3\epsilon) k_4}
\biggl({4\pi \rho_m \over 3\rho_c}
\biggr)^{1/3}F \equiv C_c \epsilon F,\label{compactness0}
\eeq
where
\beq
F \equiv 1 - {\kappa \omega^2 \over \epsilon k_2}\biggl( 
{\eta + \kappa \omega^{4/3} \over \eta - 3\kappa \omega^{4/3}}
+\epsilon \biggr) \leq 1 , \label{eqF}
\eeq
and where $C_c$ is a constant depending on $n$ (cf. Table I). 
For $n \rightarrow 3$, we find 
\beq
{M \over R} \simeq 0.445 \epsilon F. \label{compactness}
\eeq
Thus, for $\omega \rightarrow 0~(F\rightarrow 1$; infinite separation), 
we recover the well-known result $M/R=0.445\epsilon$ 
for onset of the instability against 
radial collapse \cite{chandra}. 
We note that $F$ is less than unity for finite value of $\omega$. 
Thus, the compactness of marginally stable stars against 
quasi-radial collapse in a corotating binary is smaller 
than that for spherical stars. 
This is mainly because the stellar radius is enlarged 
by the centrifugal force induced by the spin angular momentum $S$. 
Figure 1 also shows that the instability to 
quasi-radial collapse sets in for a slightly smaller central density and 
a slightly larger mass than those for spherical stars in isolation, 
for given entropy. 

The instability for a star with $\bar \rho_c < \bar \rho_1$ 
is not due to the first post-Newtonian effect; namely, 
it is not associated with quasi-radial collapse.
First, we derive the condition for the existence of such an instability. 
For stable equilibrium, the condition, 
$\pa^2 E / \pa x^2|_{M,J,K} > 0$, has to be satisfied. Equation 
(\ref{d2Ed2x}) obviously implies that the following condition 
is necessary, independent of $M$ and $x$ and irrespective of the 
first post-Newtonian term:
\beq
\epsilon k_2 -\kappa \omega^2 \biggl(
{\eta + \kappa \omega^{4/3} \over \eta - 3\kappa \omega^{4/3}}
+\epsilon \biggr) > 0. \label{limit0}
\eeq
Note that this condition is sufficient to insure stability 
in the Newtonian case.

Assuming $\omega \ll (\eta/\kappa)^{3/4}$ and $\epsilon \ll 1$, 
Eq. (\ref{limit0}) can be approximately written as 
\beqn
\omega^2 < \omega_{\rm min}^2,~~~~~
\omega_{\rm min}^2 \simeq {\epsilon k_2  \over \kappa } 
\biggl[ {\eta + \kappa(\epsilon k_2/\kappa)^{2/3} 
\over \eta -3 \kappa(\epsilon k_2/\kappa)^{2/3}} \biggr]^{-1}
\simeq 1.5 \epsilon (1 - 2.8\epsilon^{2/3}) \label{constraint}
\eeqn
or 
\beq
r \agt \biggl({8\pi \kappa \rho_m \over 3k_2\epsilon \rho_c}\biggr)^{1/3} 
\biggl[ {\eta + \kappa(\epsilon k_2/\kappa)^{2/3} 
\over \eta -3 \kappa(\epsilon k_2/\kappa)^{2/3}} \biggr]^{1/3} R 
\sim 4.6\biggl({0.001 \over \epsilon}\biggr)^{1/3} 
(1+0.9\epsilon^{2/3}) R ,\label{constr}
\eeq
where to evaluate $k_2$, $\kappa$, and $\rho_c/\rho_m$, 
we use the value for an $n=3$ Newtonian spherical polytrope (cf. Table I). 
Equation (\ref{constraint}) implies that for a very small value of 
$\epsilon$ (i.e., $\epsilon \alt 0.01$) with which 
$\omega_{\rm min}^2 \alt 0.01$, 
close orbits with $\omega^2 \agt \omega_{\rm min}^2$ 
[see Eq. (\ref{limit})] are secularly unstable; namely, 
the existence of close corotating binaries may be prohibited 
when $\epsilon$ is too small. It is noteworthy that 
all corotating binary stars with $\epsilon \rightarrow 0$ 
are secularly unstable. 

In Figs. 2 (a) and (c), we show the relations for $\bar J$ and 
$\bar\Omega (=K^{n/2}\Omega)$ as functions of $\omega^2$ 
for $\epsilon=0.003$ along a fixed value of $\bar M$ 
(here we choose $\bar M=4.05$ and 3.90). We also show the relation 
between $\bar J$ and $\bar \Omega$ near the minimum of $\bar J$ 
in Figs. 2 (b) and (d). As expected 
from Fig. 1, the curve of $\bar J$ has a minimum at 
$\omega=\omega_{\rm min}$. 
For higher mass case $\bar M=4.05$, the GR effect is important so that 
$\omega_{\rm min}^2$ is significantly less than the Newtonian 
value shown in Eq. (\ref{constraint}), i.e., $\sim 0.0042$. 
However, for $\bar M=3.90$, it approaches this value. 

Recall that the turning point theorem 
specifies nothing about the type of unstable mode or 
the growth timescale for the secular instability. 
Thus, we can suggest 
plausible outcomes. One of the most likely outcomes is that, 
given some dissipation, a binary with 
$\omega > \omega_{\rm min}$ could become secularly unstable 
to merger as in the case of stiff equation of state with 
$\Gamma \agt 2$ \cite{LRS}.  
[Namely, a corotating supermassive 
binary appears to have an innermost stable circular orbit (ISCO).] 
It should be noted that in a binary of supermassive stars, 
the system evolves as a result of thermal radiation on the 
time scale $\tau_{\rm evol}$. This implies that 
the secular instability can play a role only if the 
secular timescale is as small as $\tau_{\rm evol}$. 
As discussed in Sec. II, the dissipation timescale due to gravitational waves
$\tau_{\rm GW}$ is much longer than $\tau_{\rm evol}$. 
Hence, this secular instability is irrelevant for evolution of 
supermassive binary stars. However, viscous damping $\tau_{\rm vis}$ 
may be important. 

The minimum of $\bar J$ (i.e., ISCO) along a sequence of constant $\bar M$ 
appears before two stars come into contact 
only for restricted values of $n$ (i.e., $n \alt 1$ or $|n-3| \ll 1$). 
The reason can be explained in the following manner. 
The angular momentum of one star [cf. Eq.(\ref{eqJ})] is written in the form 
\beq
J(r)={M^{3/2}r^{1/2} \over 2\sqrt{2}} 
+ {2\sqrt{2}\kappa_n M^{3/2} R^2 \over 5r^{3/2}},\label{jnew}
\eeq
where $\kappa_n=I/(0.4MR^2) = (5\kappa/2)(4\pi \rho_m/3 \rho_c)^{2/3}$, which 
is a monotonically decreasing function of $n$ (
for Newtonian polytropes, $\kappa_0=1$, $\kappa_1 \simeq 0.653$, and 
$\kappa_3\simeq 0.188$ \cite{LRS}). 
The ISCO is located by finding zero point of $dJ/dr$ as a function of $r$, 
which is calculated as 
\beq
{dJ \over dr}={M^{3/2} \over 2^{5/2}r^{1/2}}
\biggl[1 - {6\kappa_n \over 5}\biggl({2R \over r}\biggr)^2
\biggl(1-{4r \over 3R}{dR \over dr}\biggr)\biggr]. \label{djdr}
\eeq
Here, we note that $dR/dr$ is negative, since the stellar radius 
is enlarged by the centrifugal force associated with the 
spin angular momentum which increases with decreasing $r$. 
Equation (\ref{djdr}) implies that 
the zero point of $dJ/dr$ 
for corotating binaries is produced by the effect of 
spin angular momentum and 
exists for $r \agt 2R$ if $\kappa_n$ is large enough $\agt 0.8$ 
or if $|dR/dr|$ is large enough, i.e., of $O(R/r)$, 
in the case that $\kappa_n$ is small. 
For soft equations of state with $n \sim 3$,  
$\kappa_n$ is small $\sim 0.188$, but 
$R$ rapidly increases with decreasing $r$ for close orbits. 
(Recall that for $n \sim 3$, the equatorial radius 
increases significantly even with a small magnitude of the spin angular 
momentum; e.g., for the Kepler spin velocity, 
the equator increases by a factor of 1.5 
for $n=3$, according to the Roche model \cite{ST}.) 
This rapid increase of $R$ with decreasing $r$ produces the ISCO. 

As shown in \cite{LRS,BCSST}, ISCOs also exist for sufficiently stiff 
equations of state with $n \alt 1$. In this case, $R$ does not change 
significantly with decreasing $r$ \cite{LRS}. However, $\kappa_n$ 
is of $O(1)$ for $n \alt 1$, resulting in existence of the ISCO. 
For intermediate range of $n$ with $1 \alt n \alt 3$, neither the 
magnitude of $\kappa_n$ nor the rate of increase of $R$ with decreasing 
$r$ is sufficient to produce an ISCO. 

From Fig. 2, we find that $\Omega$ has a maximum at a point where 
$\omega > \omega_{\rm min}$. The appearance of this maximum 
can be explained in the similar manner. 
Along a corotating binary sequence of fixed values of $M$ and $K$, 
$\rho_{c}$ decreases with increasing $\omega$. 
However, the orbital frequency  
$\Omega=\omega \rho_{c}^{1/2}$ does not have to increase. 
For sufficiently small $\omega$, the decreasing rate of $\rho_{c}$, 
$d\rho_{c}/d\omega$, due to increasing spin frequency of each star is 
so small that $\Omega$ increases with $\omega$. For 
close orbits, on the other hand, 
the effect of the spin becomes significant enough to enlarge 
the equator and to decrease $\rho_{c}$ by a large factor. 
Then, $\Omega$ can decrease with increasing $\omega$. 


Assuming that the timescale for secular instability, $\tau_{\rm vis}$, 
is as short as $\tau_{\rm evol}$, 
we can constrain the allowed region for existence of 
supermassive stars in corotating binary systems. For a supermassive 
star of mass $M$, $\epsilon$ can be written as 
$1.42(M/M_{\odot})^{-1/2}$ [cf. Eq. (\ref{eq142})].
Thus, Eq. (\ref{constr}) can be rewritten to
\beq
r  > 4.1R \biggl({M \over 10^6M_{\odot}}\biggr)^{1/6}. 
\eeq
On the other hand, Eq. (\ref{constr0}) has to be satisfied 
for achieving corotation. 
These two conditions constrain the allowed regions 
of $M$ and $r/R$ according to 
\beqn
&& 4.1\biggl({M \over 10^6M_{\odot}}\biggr)^{1/6} \alt {r \over R} \alt 
12 \alpha^{1/6}
\biggl({M \over 10^{6}M_{\odot}}\biggr)^{-11/24},\\
&& M \alt 6\times 10^6 \alpha^{4/15} M_{\odot}. 
\eeqn
Thus, the existence of supermassive stars in corotating binary systems 
may be strongly restricted provided the growth 
timescale of the secular instability is as short as $\tau_{\rm evol}$. 

\section{Numerical model for corotating binary}

To demonstrate that the analytical model presented in Sec. III 
is qualitatively correct, we perform numerical computation 
for equilibrium of corotating, supermassive binary stars.
To construct a numerical model, we adopt the 
conformally flat approximation of general relativity. 
In this approximation, we solve part of 
the Einstein field equations assuming that the 
spatial line element is conformally flat (see, e.g., \cite{BCSST,S} 
for equations). 
From the post-Newtonian point of view, this approximation 
is exact only up to the first post-Newtonian order. 
However, supermassive stars are not very compact, 
even at the onset of radial collapse, so that the 
higher post-Newtonian terms do not play any important role. 
Thus, the solution derived 
in this formulation provides an excellent approximation. 

To demonstrate that the results 
in Sec. III are qualitatively correct, we 
present here only one example setting $\epsilon=0.003$. The detailed 
results and detailed analysis for other parameters 
of the numerical computation will be presented elsewhere. 

The computation is performed on a $121 \times 121 \times 121$ 
Cartesian grid. We here use two numerical codes which were previously 
implemented to prepare initial conditions for mergers of 
neutron star binaries \cite{S,UE}. Computations were mainly performed 
in the code by Shibata \cite{S}, while the other code \cite{UE} 
was used to confirm the numerical results. 
The equations are solved in 
one octant ($x, y, z \geq 0$). We always assign 40 grid points 
for the major diameter of a supermassive star. 
Comparing the maximum mass for 
sufficiently distant orbits with the maximum mass for isolated 
spherical stars, 
which can be computed with high accuracy in a one dimensional code, 
it is found that with this resolution, the mass 
of the system is systematically underestimated by $\sim 0.8\%$. 
Hence, our numerical results contain an error of this order. 
To check the convergence of numerical solutions, 
we performed computations on a $91 \times 91 \times 91$ 
grid, covering the major diameter of the star by 30 grid points. 
In this case, the mass is systematically 
underestimated by $\sim 1.5\%$, indicating 
that convergence with increasing the resolution is achieved. 

In Fig. 3, we show the dependence of 
$\bar M_*$ versus $\bar \rho_c$ (compare with Fig. 1). 
Here, $M_*$ denotes the rest mass of one star. 
The solid lines denote 
the relation between $\bar M_*$ and $\bar \rho_c$ for fixed values of 
$\hat d \equiv (r_{\rm out}+r_{\rm in})/(r_{\rm out}-r_{\rm in})=1.3$, 
1.5, and 3.05 
where $r_{\rm out}$ and $r_{\rm in}$ denote the coordinate 
distance from the origin to the outer and inner edges of star 
along the axis which connects the centers of mass of the two stars.
$\omega^2(=\Omega^2/\rho_c)$ is 0.006, 0.0046, and 0.00068 
for $\hat d=1.3$, 1.5 and 3.05, respectively. 
The dashed curves show the relation of $\bar M_*$ versus $\bar \rho_c$ 
for constant values of $\bar J$. 

As expected from the analytical results 
for $\bar J \agt 410$, the curve for constant $\bar J$ has 
one minimum at $\bar \rho_1$ and one maximum at 
$\bar \rho_2 > \bar \rho_1$: The former is associated 
with the ISCO and the latter with the 
onset of quasi-radial collapse. 
It is found that the turning point at $\bar \rho_1$ approaches 
to the curve of $\omega^2 \sim 0.0042$ for the Newtonian limit 
($\bar \rho_1 \rightarrow 0$), as indicated by the analytical 
study [cf. Eq. (\ref{constraint})]. 

In Fig. 4, we show the relation between $\bar J$ and $\bar \Omega$ 
for $\bar M_*=4.04$ near the minimum of $\bar J$ [compare with Fig. 2 (b)]. 
It is seen that the minimum of $\bar J$ and the maximum of 
$\bar \Omega$ exist, illustrating that 
the analytical results presented in Sec. III are qualitatively correct and 
quantitatively fairly accurate. All these results 
demonstrate the robustness of analytical modeling for corotating, 
supermassive binary stars.

\section{Quasi-stationary evolution of supermassive binary stars
}

In this section we determine evolutionary sequences of corotating, 
supermassive binary stars to the onset of instability or merger. 
As discussed in Sec. II, supermassive stars 
dissipate energy by thermal radiation. 
We assume that rest mass is conserved during the evolution, 
since spin is far below Kepler limit 
in corotating systems. 
Although a small amount of mass of system may be ejected due to 
a stellar wind in reality \cite{ZN}, 
we neglect such effects for simplicity. 
Since the timescale of gravitational radiation reaction 
is too long to dissipate the angular momentum and 
because of mass conservation, we 
assume that the angular momentum of the system is also conserved. 
We also assume that the effective 
adiabatic constant $\Gamma$ is unchanged during the evolution 
for simplicity. 
Strictly speaking, this assumption is not valid since 
$\beta$ depends on the entropy of the radiation which 
decreases during the evolution. This treatment is adequate 
only when the fractional 
decrease in the entropy is not very large. 

In this setting, the energy $E$ for 
an equilibrium star is a function of $K$ and $x$, i.e., $E(K,x)$, and 
the evolution equation for $E$ can be written as
\beq
\dot E = k_1 \dot K M x^{3/n} + {\pa E \over \pa x}\Big|_{M,J,K} \dot x
=k_1 \dot K M x^{3/n},
\eeq
where ``$~\dot{}~$'' denotes the time derivative, and 
where $(\pa E/\pa x)|_{M,J,K}=0$ since we assume quasi-stationary  
evolution along an equilibrium track. From this equation, 
it is immediately found that $K$ decreases 
with evolution since $\dot E$ is negative. 

The decrease of $K$ implies that $\bar M$ and $\bar J$ increase. 
However, $\bar J/\bar M^2=J/M^2$ remains constant. 
Thus, the evolutionary sequence is specified by  
the value of $\bar J/\bar M^2$.
In Fig. 5, we show $\bar M$ as a function of $\bar \rho_c$ for 
a fixed value of $\bar J/ \bar M^2(=28, 31$ and 35) and $\epsilon=0.003$ 
as an example (thick solid lines). 
Along these curves, $\bar M$ has to increase. 
This implies that 
if a corotating binary initially resides 
in the right-hand side of the dotted curve (which indicates 
the turning points), $\bar \rho_c$ increases with decreasing 
$\omega$. If a corotating binary resides in the left-hand side of 
the dotted curve, $\bar \rho_c$ decreases with increasing $\omega$. 
Since we assume that $J$ is conserved, 
$\Omega$ should slightly decrease (increase) with decreasing 
(increasing) $\omega$ in these evolutionary sequences 
[cf. Eq. (\ref{dwdx})]. Therefore, if a corotating binary 
initially resides in the right-hand (left-hand) side of the 
dotted curve, the orbital separation $r$ slightly increases (decreases) 
with evolution.

As we found in Sec. III, 
supermassive stars on the left-hand side of the 
dotted line are secularly unstable to merger. 
Thus, if this instability grows in a timescale 
as short as $\tau_{\rm evol}$, 
the final fate is not determined by thermal radiation but by the 
secular dissipation timescale \footnote[2]{
A point of dynamical instability is likely to arise beyond 
the point of secular instability, but this point cannot be 
identified by the present treatment 
that assumes corotation (see, e.g., \cite{LRS})}. 
Only when the system evolves by means of thermal emission, 
supermassive stars quasi-stationarily evolve with decreasing central density 
and increasing $\omega$ to merger. 
In either case, the final outcome will likely 
be a single, rotating supermassive star with a 
surrounding disk as a result of merger. 
Such a rotating star will presumably be rapidly rotating, so that 
its subsequent evolution may be essentially the same 
as for a single isolated rotating star as discussed in \cite{BS}.

In the case when the binary resides in the right-hand side of 
the dotted line, the evolution will proceed via 
one of several scenarios. 
If the viscous timescale with regard to the tidal torque is 
shorter than the thermal evolution timescale throughout the 
entire evolution, corotation is preserved, and 
the sequence terminates at the point of 
unstable, quasi-radial collapse. We refer to this scenario 
as scenario (A). 
However, viscosity may not be strong 
enough to preserve corotating orbits for small 
$\omega$, i.e., $\omega < \omega_{\rm vis}$, 
where [cf. Eqs. (\ref{constrr}) and (\ref{limit})], 
\beq
\omega_{\rm vis}^2 \sim 5.6 \times 10^{-5} \alpha^{-1/3} 
\biggl({ M/R \over 10^{-3} }\biggr)^{-5/4} 
\biggl({ M \over 10^6M_{\odot}}\biggr)^{3/4}
\agt 10^{-4} \alpha^{-1/3} 
\biggl({ M \over 10^6M_{\odot}}\biggr)^{11/8}. 
\eeq
Here we use Eq. (\ref{compactconstr}) to derive the 
second inequality. 
In this case, the orbit may depart from corotation 
before reaching the turning point. We refer to this scenario as 
scenario (B).

Scenario (A) is possible only for relatively small mass 
since the condition $\omega_{\rm crit} \geq \omega_{\rm vis}$ 
is necessary. This condition reduces to 
\beqn
&&{M \over R} \agt 2 \times 10^{-4} \alpha^{-4/15}
\biggl({M \over 10^6 M_{\odot}}\biggr), \\ 
&&M \alt 4\times 10^6 \alpha^{4/15}M_{\odot}.
\eeqn
Thus, only for supermassive binary stars of fairly small mass and 
of fairly large compactness, 
corotation can be preserved until the onset of 
quasi-radial instability.

Assuming that corotation is preserved 
through the entire evolution [i.e., scenario (A)], 
we estimate the non-dimensional, 
spin parameter of each star, $q \equiv S/M^2$, at the 
onset of quasi-radial collapse 
in order to assess the final remnant following the collapse. 
Since the star in corotation has a spin 
angular velocity equal to $\Omega$, $q$ is 
calculated according to 
\beq
q={I\Omega \over M^2}
=\kappa \biggl({3\rho_c \over  4\pi \rho_m}\biggr)^{-1/6}
\biggl({R \over M}\biggr)^{1/2}\omega. 
\eeq
Using Eq. (\ref{compactness0}), we 
find that $q$ at the onset of 
quasi-radial collapse for $\epsilon \ll 1$ is 
\beq
q = q_0 \omega \epsilon^{-1/2}F^{-1/2},\label{qeq}
\eeq
where 
\beq
q_0 \equiv \kappa \biggl({2k_4(1-3\epsilon) \over 3k_2}\biggr)^{1/2};
\eeq
values of $q_0$ are at most $\sim 0.4$ (cf. Table I). 
As mentioned in Sec. III, $\omega \leq \omega_{\rm crit}$ at a 
turning point for quasi-radial collapse 
(e.g., the thick dotted line in Fig. 5). 
With this constraint and Eq. (\ref{newcon}), 
Eq. (\ref{qeq}) reduces to 
\beq
q < 0.6 q_0 F^{-1/2} \sim 0.25 F^{-1/2}. 
\eeq 
Since $\omega$ is less than $\omega_{\rm crit}$ 
[cf. Eq. (\ref{newcon})] at 
the onset of quasi-radial collapse, $F \agt 0.5$ [cf. Eq. (\ref{eqF})]. 
Therefore, if quasi-radial instability sets in 
for supermassive stars in corotating binaries, 
$q$ has to be much smaller than unity. Consequently, 
the resulting spin is below the Kerr limit ($q=1$) and 
the final fate is likely to be a slowly rotating black hole. 
This result is due to the spin angular momentum being effectively 
transferred to orbital angular momentum 
via viscosity during the evolution.

For scenario (B), in which the corotation is not preserved for 
$\omega < \omega_{\rm vis}$, we can consider two scenarios 
for the subsequent evolution. 
In one scenario, each star reaches a mass-shedding limit 
as a result of the decrease of stellar radius. 
It then evolves along a mass-shedding sequence ejecting 
mass from the equator until the onset of 
quasi-radial instability \cite{BS}. 
We refer to this scenario as scenario (Ba). 
As discussed in \cite{BS},  
the final fate of each star after the onset of the 
instability is not clear, since 
$q$ is nearly equal to unity at the onset of the instability, 
and rotation may inhibit direct collapse to a black hole. 
In the other scenario, 
each star in the binary shrinks, conserving spin angular momentum 
through the entire evolution, until quasi-radial collapse. 
This is the case in which 
the stellar radius at $\omega=\omega_{\rm vis}$ is sufficiently 
small to reach the turning points before reaching 
a mass-shedding limit. 
We refer to this scenario as scenario (Bb). In this case, $q$ 
parameter could be smaller than unity so that 
the final outcome is likely to be a black hole of 
moderately large spin. 

We derive the criterion for the stellar 
radius at $\omega=\omega_{\rm vis}$ to follow scenario (Bb). 
From Eq. (\ref{constrr}), the angular velocity of each star 
at $\omega=\omega_{\rm vis}$ is 
\beq
\Omega_{0}^2 \sim 7 \times 10^{-4} \alpha^{-1/2}{M \over R_0^3}
\biggl({M/R_0 \over 10^{-3}}\biggr)^{-5/4}
\biggl({M \over 10^{6}M_{\odot}}\biggr)^{3/4},
\eeq
where $R_0$ is the stellar radius at $\omega=\omega_{\rm vis}$. 
With decreasing stellar radius $R < R_0$, $\omega$ 
becomes smaller than $\omega_{\rm vis}$. 
For $\omega < \omega_{\rm vis}$, the spin angular momentum 
of each star remains constant until the star reaches a 
mass-shedding limit at which the angular velocity 
approximately becomes \cite{ZN,ST,BS}
\beq
\Omega_{\rm shed}^2 \simeq {8M \over 27R^3}.
\eeq
Because of the conservation of the spin angular momentum and mass, 
the relation, $R^2\Omega = R_0^2 \Omega_0$, approximately 
holds. Hence, 
\beq
\Omega^2 \sim 7\times 10^{-4} \alpha^{-1/2}{M R_0 \over R^4}
\biggl({M/R_0 \over 10^{-3}}\biggr)^{-5/4}
\biggl({M \over 10^{6}M_{\odot}}\biggr)^{3/4}. 
\eeq
For the second scenario, $\Omega$ has to be less than 
$\Omega_{\rm shed}$ at the onset of quasi-radial collapse, i.e., 
\beq
{R \over R_0 } \agt 0.07 \alpha^{-2/9}
\biggl({M/R \over 10^{-3}}\biggr)^{-5/9}
\biggl({M \over 10^{6}M_{\odot}}\biggr)^{1/3}. 
\eeq
Thus, $R_0$ has to be less than $\sim 10$ times of 
stellar radius at the onset of quasi-radial collapse 
for $M \sim 10^6M_{\odot}$. 

In Fig. 6, we present a schematic diagram for the evolution of 
supermassive binaries of $M=10^6M_{\odot}$ for $\alpha=1$. 
The horizontal and 
vertical axes denote $M/R$ and $r/R$, respectively. 
Since $M$ is conserved and $r$ remains approximately constant 
in negligible gravitational wave emission, 
evolution proceeds at fixed $M/r$ approximately 
(see solid and dashed lines). 
A binary with $r/R < 2$ (below the lower long-dashed line) 
should be regarded as a single star, so that 
only the region with $r/R > 2$ is relevant here. 
Slowly rotating supermassive stars 
with $M/R \agt  6.3\times 10^{-4}$ and $S/M^2 \ll 1$ 
are unstable against quasi-radial collapse 
(the right-hand side of the vertical long-dashed line). 
A binary is corotating if it resides in 
the region surrounded by three long-dashed lines. 
A corotating binary with $r/R \agt 4.1$ 
(above the dotted line) is secularly stable, so that 
if a corotating binary resides above (below) the dotted line, 
$r/R$ increases (decreases) with evolution. 
(We note that for non-corotating binaries in which tidal, viscous 
effect is negligible, such secular instability would not exist, 
implying that $r/R$ would always increase for them.) 

From these results, we find the following conclusion for $M=10^6M_{\odot}$:
(1) If $M/R$ of a corotating binary is initially larger than 
$\sim 2 \times 10^{-4}$, the corotation is preserved 
until each supermassive star in the binary reaches the onset of 
quasi-radial collapse [scenario (A)]. 
In this case, the product is a binary black hole of small spin 
parameter. 
(2) If $M/R$ of a corotating binary is initially larger than 
$\sim 6 \times 10^{-5}$, each supermassive star in the binary 
reaches the onset of quasi-radial collapse before 
reaching the mass-shedding limit [scenario (Bb)]. 
In this case, the product is a binary black hole of moderately large 
spin. 
(3) If $M/R$ is initially smaller than $\sim 6\times 10^{-5}$, 
stable, supermassive binary stars in corotation 
can never be achieved [scenario (Ba)]. 
In this case, the outcome is not clear. If the supermassive 
stars directly collapse to black holes conserving the spin and mass, 
the product would be a binary black hole of large spin $q \sim 1$.  

\section{Implication for gravitational wave detection}

In this section, we estimate the frequency and amplitude of 
gravitational waves from supermassive binary stars in stable 
circular orbits and from 
collapse to a black hole [i.e. for scenario (A) and (Bb)]. 
Possibilities for generating burst and quasi-periodic 
gravitational waves after the onset of 
quasi-radial instability in scenario (Ba) are the same as 
those discussed in \cite{BS}. 

Frequency $f$ and amplitude $h$ of gravitational waves 
from supermassive binary stars in circular orbits are calculated as
\beqn
&&f \sim 3\times 10^{-6} {\rm Hz} \biggl({1000M \over r}\biggr)^{3/2}
\biggl({10^6 M_{\odot} \over M}\biggr),\label{frequency} \\
&&h \sim 10^{-19} \biggl({1000M \over r}\biggr)
\biggl({M \over 10^{6}M_{\odot}}\biggr)
\biggl({3000{\rm Mpc} \over D}\biggr), \label{amplitude}
\eeqn
where $D$ denotes the distance from the source 
to detectors. Since $r > 1000M$ for supermassive binary stars 
with $M > 4\times 10^5 M_{\odot}$, 
the frequency is smaller than $\sim 3 \times 10^{-6}$ 
which is too low to be detected by LISA \cite{KIP0,LISA}. 

Close binaries will be secularly unstable to merger to a single rotating 
star. During merger, the coalesced object is likely to 
form a stable ellipsoid which could emit quasi-periodic 
gravitational waves. Even in this case, the frequency will be 
similar to that given in Eq. (\ref{frequency}) and 
is still too low, since the radius of the ellipsoid is likely to be 
larger than $\sim 1000M$. 

Supermassive stars in a stable binary orbit likely collapse 
to black holes after quasi-stationary contraction but prior to merger. 
For the case 
when the orbital separation and compactness of each star 
are initially small, the black holes are directly formed after 
onset of quasi-radial instability according to scenarios (A) and (Bb). 
As we found in Sec. V, supermassive stars  are 
rotating with $0.3 \alt q < 1$ at the onset of quasi-radial 
instability. Since the 
collapse is non-spherical, gravitational waves are emitted. 
Gravitational waves during formation of black holes are likely 
dominated by quasi-normal modes of black holes \cite{SP}. 
Perturbation studies for 
the frequency of quadrupole quasi-normal modes provide \cite{Leaver}
\beq
f \sim (0.03-0.08)M^{-1} \simeq (6-15)\times 10^{-3} 
\biggl({M \over 10^{6}M_{\odot}}\biggr)^{-1} {\rm Hz},~~~~~~
{\rm for}~0 \leq q < 1.
\eeq
Thus, the frequency for $M \sim 10^6M_{\odot}$ 
is in the range in which LISA is expected to be most sensitive. 
The strength of the gravitational wave signal likely depends strongly 
on $q$ and on the dynamics of collapse, so that it is difficult to assess 
without numerical simulation. An order-estimate yields \cite{KIP} 
\beq
h \sim 10^{-19} \biggl({\varepsilon \over 10^{-6}}\biggr)^{1/2}
\biggl({10^{-3}{\rm Hz} \over f}\biggr)^{1/2}
\biggl({3000{\rm Mpc} \over D}\biggr)
\biggl({M \over 10^6M_{\odot}}\biggr)^{1/2},
\eeq
where $\varepsilon$ denotes the ratio of the total emitted energy 
of gravitational waves to the gravitational mass energy $M$. 
Thus, as long as $\varepsilon \agt 10^{-10}$, 
detection of gravitational waves will be possible by LISA 
in which the sensitivity around $f \sim 10^{-3}$Hz is $\sim 10^{-21}$. 
A numerical relativistic work \cite{SP} indicates that 
$\varepsilon$ depends strongly on $q$ and initial conditions. 
Since the simulations in \cite{SP} are carried out only for collapse of 
neutron stars, it is necessary to perform a simulation 
preparing a realistic initial condition and using 
realistic soft equations of state 
for supermassive stars to clarify the magnitude of $\varepsilon$. 
However, Ref. \cite{SP} reports that as long as $q \agt 0.2$, 
$\varepsilon$ could be larger than $10^{-6}$. If this relation would 
hold for collapse of supermassive stars, future detection of 
gravitational waves for scenarios (Ba) and (Bb) would be possible. 

If supermassive stars eventually collapse to a black hole, 
a black hole binary will be formed and 
subsequently decreases orbital separation radiating 
gravitational waves during inspiraling in a timescale $\tau_{\rm GW}$ 
[see Eq. (\ref{taugw})]. The frequency and amplitude are given by 
Eqs. (\ref{frequency}) and (\ref{amplitude}), which 
show that supermassive binary black holes with 
$r \alt 200M(M/10^6M_{\odot})^{-3/2}$ emit gravitational waves 
of $f \agt 10^{-4}$Hz and can be strong sources of LISA.

\section{Summary}

We have investigated equilibrium and stability of supermassive 
stars in two-body binary systems. 
We found that supermassive binary stars in close, corotating 
orbits with $r/R \alt 4(M/10^6M_{\odot})^{1/6}$ are secularly unstable. 
We also have shown that the corotation cannot 
be achieved for distant orbits with 
$r/R \agt 12(M/10^6M_{\odot})^{-11/24}$, 
since the timescale for viscous transfer of the angular momentum 
is longer than the evolution timescale due to radiation of 
thermal energy. These facts suggest that 
existence of stable supermassive stars in corotating binaries 
is highly restricted. In particular for 
massive systems $M \agt 6 \times 10^6M_{\odot}$, corotation 
cannot be achieved. Evolution in such a massive binary 
does not experience 
transfer between orbital and spin angular momentum. 
Namely, the evolution of the velocity field of one star 
is not affected by its companion star. In this case, each supermassive 
star will be rapidly and rigidly rotating, as in the case of 
isolated supermassive stars. 

As discussed in \cite{BS}, rotating supermassive stars in isolation 
will evolve along a mass-shedding sequence 
to quasi-radial collapse. At the onset of the instability, 
the nondimensional spin parameter $q$ is close to unity, 
suggesting that the collapsing star may not form a black hole 
in a straightforward manner. 
During the collapse, the centrifugal force may inhibit 
the collapse to a black hole near the equatorial plane to form a disk. 
After the formation of disk, non-axisymmetric instabilities 
may play an important role in forming a bar, or in fissioning to
several blobs. On the other hand, if supermassive stars 
collapse directly to black holes, the resulting black holes will have 
a large spin near Kerr limit $q \sim 1$. 
Only numerical simulation in full general relativity 
can distinguish these fates.

For supermassive stars in binaries of fairly small mass 
$\alt 4 \times 10^6M_{\odot}$, 
small initial orbital separation $r\sim 4R$, and 
initially small compactness, corotating orbits persist 
until the onset of quasi-radial collapse. In this case, the $q$ parameter 
of each star at the onset of the instability is much less than unity 
since spin angular momentum can be transferred to orbital 
angular momentum by viscosity during the evolution. 
The final fate of the system is likely to be a binary black hole 
of small spin, 
although it is achieved for restricted initial conditions. 

\acknowledgments

We thank T. Baumgarte for helpful discussion and comments. 
M.S. thanks the Department of Physics of the University of Illinois 
at Urbana-Champaign (UIUC), 
where most of this work was done, for warm hospitality 
and JSPS for support. Numerical computations were performed on the 
FACOM VX/4R machines in the data processing center of the 
National Astronomical Observatory of Japan. 
This work was supported by NASA grant NAG 5-7152 at UIUC.

\appendix

\section{Roche binary}

In this appendix, 
we briefly investigate equilibrium and stability of 
supermassive stars in the Roche binary. In this problem, 
we treat binaries composed of a point body of mass 
$m$ and a supermassive star of mass $M$. We again assume that 
the equation of state for the supermassive star is 
given by Eq. (\ref{pressure}). 
In this setting, 
the total energy of the system $E$ is written in the form 
\beqn
E=k_1KM x^{3/n} - k_2M^{5/3}x - k_4M^{7/3}x^2 -{Mm \over r}
  +{I \over 2}\Omega^2 +{Mm \over 2(M+m)}r^2\Omega^2,
\eeqn
where $\Omega$ here is determined by 
\beq
\Omega=\sqrt{{M+m \over r^3}},
\eeq
and other notations are the same as those in Sec. III. 
The total angular momentum is written
\beq
J={Mm \over M+m}r^2\Omega + I\Omega
=(\eta'\Omega^{-1/3}+\kappa x^{-2}\Omega)M^{5/3},
\eeq
where $\eta'\equiv m/(M+m)^{1/3}M^{2/3}$. 
Taking the first derivative of $E$ with respect to $x$ 
fixing $M$, $J$ and $K$ yields a condition for equilibrium 
\beqn
0={\pa E \over \pa x}\Big|_{M,J,K}={3k_1 \over n}K M x^{3\epsilon} 
- k_2 M^{5/3}-2k_4 M^{7/3}x 
+{\kappa \Omega^2 M^{5/3} \over x^3},\label{dedx1}
\eeqn
where we use 
\beq
{\pa \Omega \over \pa x}\Big|_{M,J}={2\kappa \over x^3}\Omega\biggl(
-{\eta' \over 3} \Omega^{-4/3}+{\kappa \over x^2}\biggr)^{-1}.
\eeq
Equation (\ref{dedx1}) is completely the same as Eq. (\ref{dedx}), 
implying that we can construct equilibria in the same manner 
as that in Sec. III. 

The second derivative of $E$ with respect to $x$ becomes 
\beqn
{\pa^2 E \over \pa x^2}\Big|_{M,J,K}
=&&{9\epsilon k_1 \over n}K M x^{3\epsilon - 1}
-2k_4M^{7/3}
-{3\kappa M^{5/3}\Omega^2 \over x^{4}}
\biggl({\eta' x^2 + \kappa\Omega^{4/3} \over \eta' x^2 -3\kappa\Omega^{4/3}}
\biggr) \nonumber \\
=&&
3\epsilon k_2 M^{5/3}x^{-1} -2(1-3\epsilon)k_4M^{7/3}
-{3\kappa \Omega^2 M^{5/3} \over x^{4}} \biggl(
{\eta' x^2 + \kappa \Omega^{4/3} \over \eta' x^2 - 3\kappa \Omega^{4/3}}
+\epsilon \biggr).
\eeqn
Thus, the stability can be investigated 
simply by exchanging $\eta$ in Eq. (\ref{d2Ed2x}) 
to $\eta'$, so that the qualitative feature on the stability is 
essentially the same as that shown in Sec. III. 
A sufficient condition for existence of a stable 
equilibrium in close orbits ($\pa^2 E/\pa x^2|_{M,J,K} > 0$) is given by
\beq
\epsilon k_2 
-\kappa \omega^2 \biggl(
{\eta' + \kappa \omega^{4/3} \over \eta' - 3\kappa \omega^{4/3}}
+\epsilon \biggr) > 0, \label{A7}
\eeq
where 
\beq
\omega^2\equiv {\Omega^2 \over x^3}=
{4\pi \over 3}{\rho_m \over \rho_c}\biggl(1+{m \over M}\biggr)
\biggl({R \over r}\biggr)^3. 
\eeq
Equation (\ref{A7}) is approximately written as 
\beq
\omega^2 < {\epsilon k_2 \over \kappa}.\label{seigen}
\eeq
Thus, we find that close orbits with $r \sim R$ are secularly 
unstable for a small 
value of $\epsilon$, i.e., $\epsilon \alt 0.05(1+m/M)$, in 
Roche binary. 

\section{Effect of the tidal energy}

In this section, we briefly review the effect of tidal energy 
to the equilibrium and stability of supermassive stars in binary systems. 
In the following, we consider supermassive binary stars of equal 
mass and of no spin for simplicity. 

Including the tidal energy but neglecting the 
effect of the spin, the energy of one star is written as \cite{LAI} 
\beqn
E=k_1KM x^{3/n} - k_2M^{5/3}x - k_4M^{7/3}x^2 
  +{M \over 8}r^2\Omega^2 -{M^2 \over 2r}
-{\lambda M^{11/3} \over r^6 x^5},
\eeqn
where $\Omega$ is determined from $\pa E/\pa r|_{M,J,K}=0$ as 
\beq
\Omega=\sqrt{{2M \over r^3}+{24\lambda M^{8/3} \over r^8x^5}},
\eeq
and 
\beq
\lambda={75 \over 16}\kappa^2 \biggl(1- {n \over 5}\biggr)
\biggl({3 \rho_c \over 4\pi \rho_m}\biggr)^{1/3}.
\eeq
Our notation is the same as that in Sec. III, except that 
the angular momentum of one star is written as 
\beq
J={Mr^2 \over 4}\Omega. 
\eeq

Taking the first derivative of $E$ with respect to $x$ 
fixing $M$, $J$ and $K$ yields a condition for equilibrium 
\beqn
0={\pa E \over \pa x}\Big|_{M,J,K}={3k_1 \over n}K M x^{3\epsilon} 
- k_2 M^{5/3}-2k_4 M^{7/3}x 
+{5\lambda M^{5/3}\over 64}
\biggl({3 \rho_c \over 4\pi \rho_m}\biggr)^{-2}
\biggl({2R \over r}\biggr)^6, \label{dedt3}
\eeqn
where we use 
\beq
{\pa r \over \pa x}\Big|_{M,J}={60\lambda M^{5/3} \over r^4 x^6}
\biggl(1-{48\lambda M^{5/3} \over x^5 r^5}\biggr)^{-1}. 
\eeq
For $n=3$, $\lambda \simeq 0.76$, and the fourth term of 
Eq. (\ref{dedt3}) is written for $n=3$ as 
$3.5\times 10^{-4}M^{5/3}(2R/r)^6$. Comparing with Eq. (\ref{dedx}), 
it is found that the magnitude of the tidal energy is by a factor 
$\sim 0.02(2R/r)^3$ smaller than the magnitude of the spin effect 
for corotating binaries, implying that 
the effect of the tidal energy for 
equilibria of supermassive stars is much less important.

The second derivative of $E$ with respect to $x$ becomes 
\beqn
{\pa^2 E \over \pa x^2}\Big|_{M,J,K}
=&&{9\epsilon k_1 \over n}K M x^{3\epsilon - 1}
-2k_4M^{7/3}
-{30\lambda M^{11/3}\over r^6 x^7}
{x^5 r^5 + 12\lambda M^{5/3} \over x^5 r^5 - 48\lambda M^{5/3}}
\nonumber \\
=&&
3\epsilon k_2 M^{5/3}x^{-1} -2(1-3\epsilon)k_4M^{7/3}
-{15 \lambda M^{5/3}\over 32 x}\biggl({2R \over r}\biggr)^6
\biggl({3 \rho_c \over 4\pi \rho_m}\biggr)^{-2}
\biggl(
{x^5 r^5 + 12\lambda M^{5/3} \over x^5 r^5 - 48\lambda M^{5/3}}
+{\epsilon \over 2}\biggr).
\eeqn
Thus, the condition for existence of stable supermassive stars 
is approximately written as
\beq
\epsilon > {5 \lambda \over 32}
\biggl({3 \rho_c \over 4\pi \rho_m}\biggr)^{-2}
\biggl({2R \over r}\biggr)^6 \sim 7 \times 10^{-4} 
\biggl({2R \over r}\biggr)^6.
\eeq
In determining the stability of the non-spinning binary system, 
the tidal energy can important 
for very small $\epsilon$ as $\epsilon < 10^{-3}$ and 
for very close orbits with $r \sim 2R$. 
However, as long as $\epsilon \agt 7\times 10^{-4}$, it is not 
important at all.

\noindent 
\begin{table} 
\caption{Nondimensional structure constant for Newtonian spherical polytrope 
with $n \simeq 3$. $\epsilon$ denotes $1/n-1/3$. 
}
\end{table}

\vskip 5mm
\noindent
\begin{center}
\begin{tabular}{|c|c|c|c|c|c|c|c|c|c|} \hline
\hspace{3mm} $\epsilon$ \hspace{3mm} &
\hspace{3mm} $\xi_1$ \hspace{3mm} &
\hspace{2mm} $|\theta(\xi_1)|$ \hspace{2mm} &
\hspace{1mm} $k_1$ \hspace{1mm} & 
\hspace{1mm} $k_2$  \hspace{1mm} &
\hspace{0mm} $k_4$    \hspace{0mm} & 
\hspace{1mm} $\kappa$   \hspace{1mm} &
\hspace{0mm} $\rho_c/\rho_m$ \hspace{0mm} & 
\hspace{0mm} $C_c$ \hspace{0mm} & 
\hspace{0mm} $q_0$ \hspace{0mm}  \\ \hline 
0      & 6.89685 & 0.0424298 & 1.75579 & 0.639000 & 0.918294 & 0.415248 
& 54.1825  & 0.44465 & 0.4065 \\ \hline 
0.0001 & 6.89334 & 0.0424785 & 1.75520 & 0.639006 & 0.918215 & 0.415118
& 54.0928  & 0.44512 & 0.4062 \\ \hline
0.001  & 6.86202 & 0.0429176 & 1.74992 & 0.639654 & 0.917504 & 0.413954 
& 53.2960  & 0.44929 & 0.4042 \\ \hline 
0.003  & 6.79394 & 0.0438962 & 1.73829 & 0.640950 & 0.915917 & 0.411410
& 51.5909  & 0.45866 & 0.3997 \\ \hline
0.005  & 6.72789 & 0.0448786 & 1.72679 & 0.642233 & 0.914321 & 0.408923
& 49.9711  & 0.46814 & 0.3954 \\ \hline
\end{tabular}
\end{center}

\newpage

\begin{figure}[t]
\vspace*{0mm}
\begin{center}
\epsfxsize=3.5in
~\epsffile{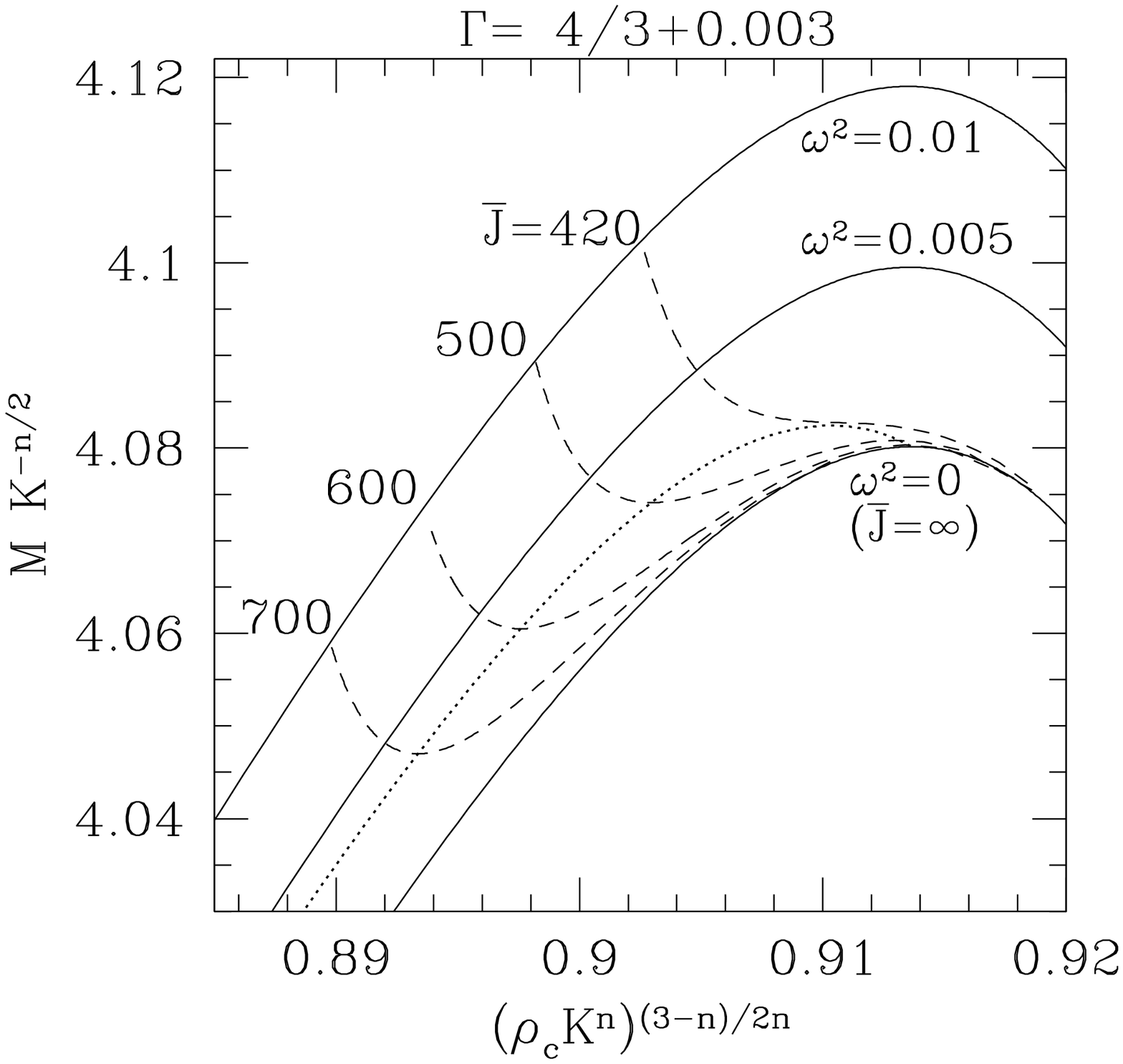}
\end{center}
\vspace*{-2mm}
\caption{Analytic mass-density relation for $\epsilon=0.003$. 
The dependence of $\bar M$ on $\bar \rho_{c}^{(3-n)/2n}$ for 
fixed values of $\omega^2$ is indicated by the solid lines. 
The dependence of $\bar M$ on $\bar \rho_{c}^{(3-n)/2n}$ for 
fixed values of $\bar J$ is shown by the dashed lines. 
The dotted line indicates the sequence of the turning points. 
}
\end{figure}

\begin{figure}[t]
\vspace*{0mm}
\begin{center}
\epsfxsize=3.5in
(a)~\epsffile{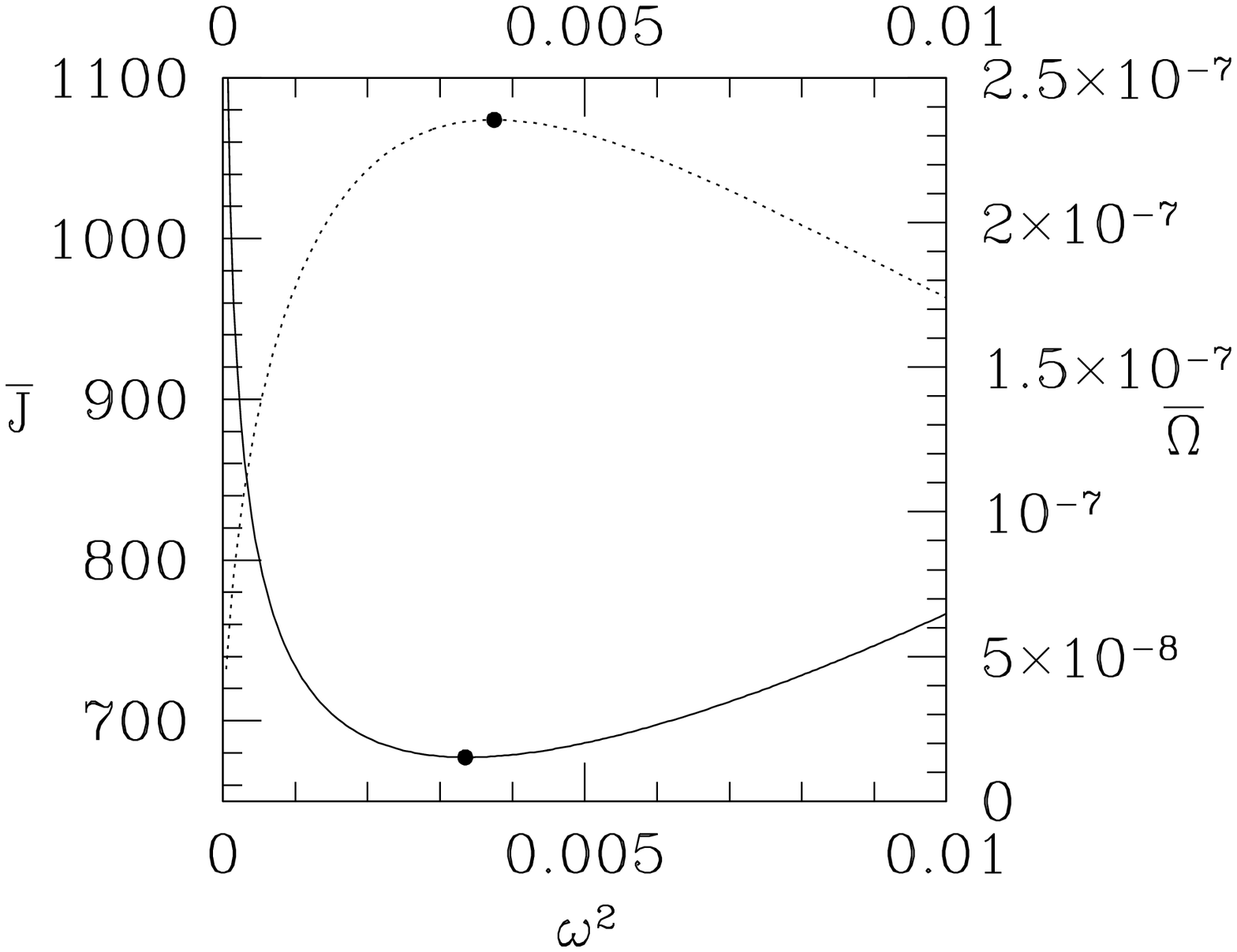}
\epsfxsize=3.5in
(b)~\epsffile{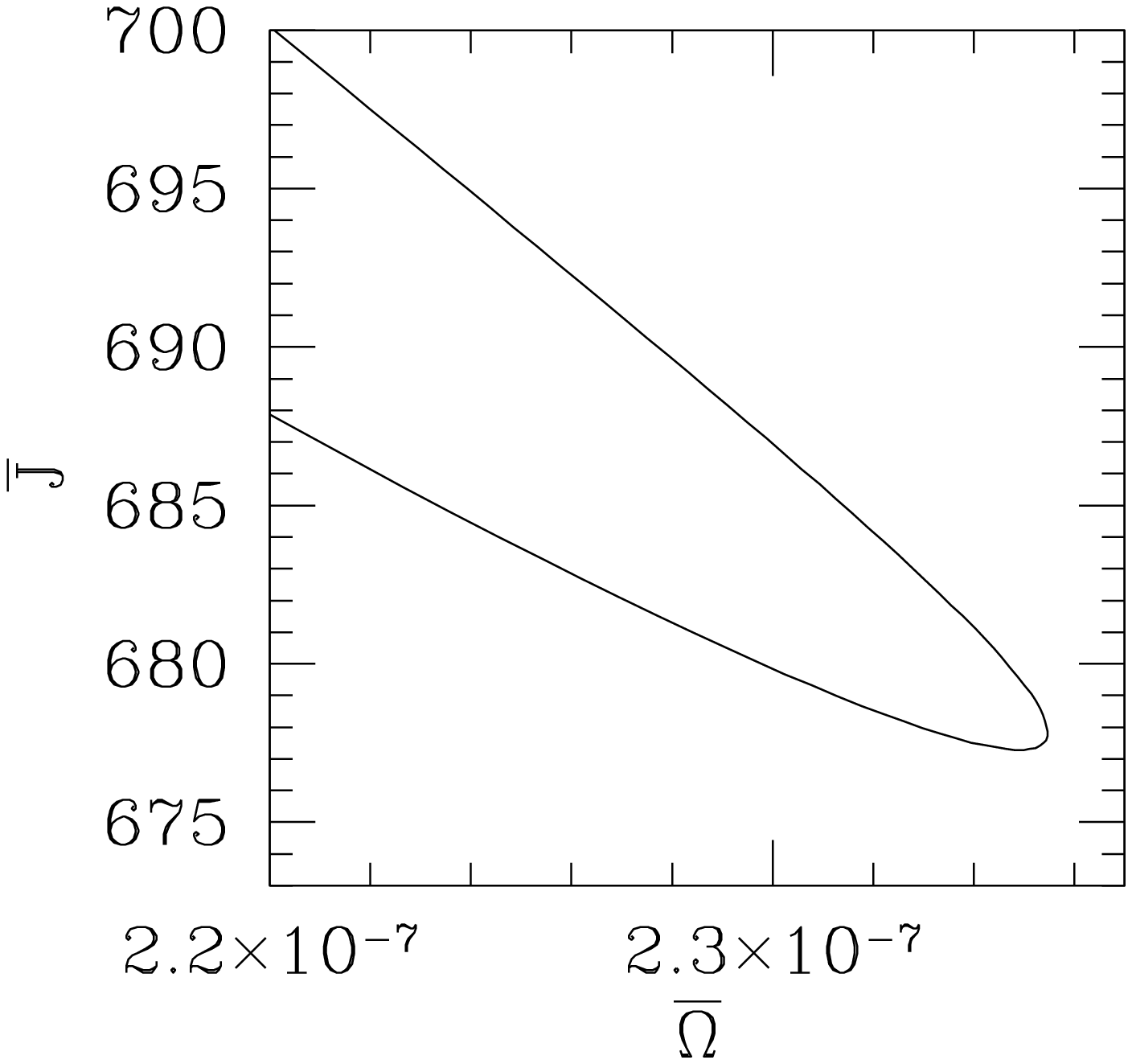}\\
\epsfxsize=3.5in
(c)~\epsffile{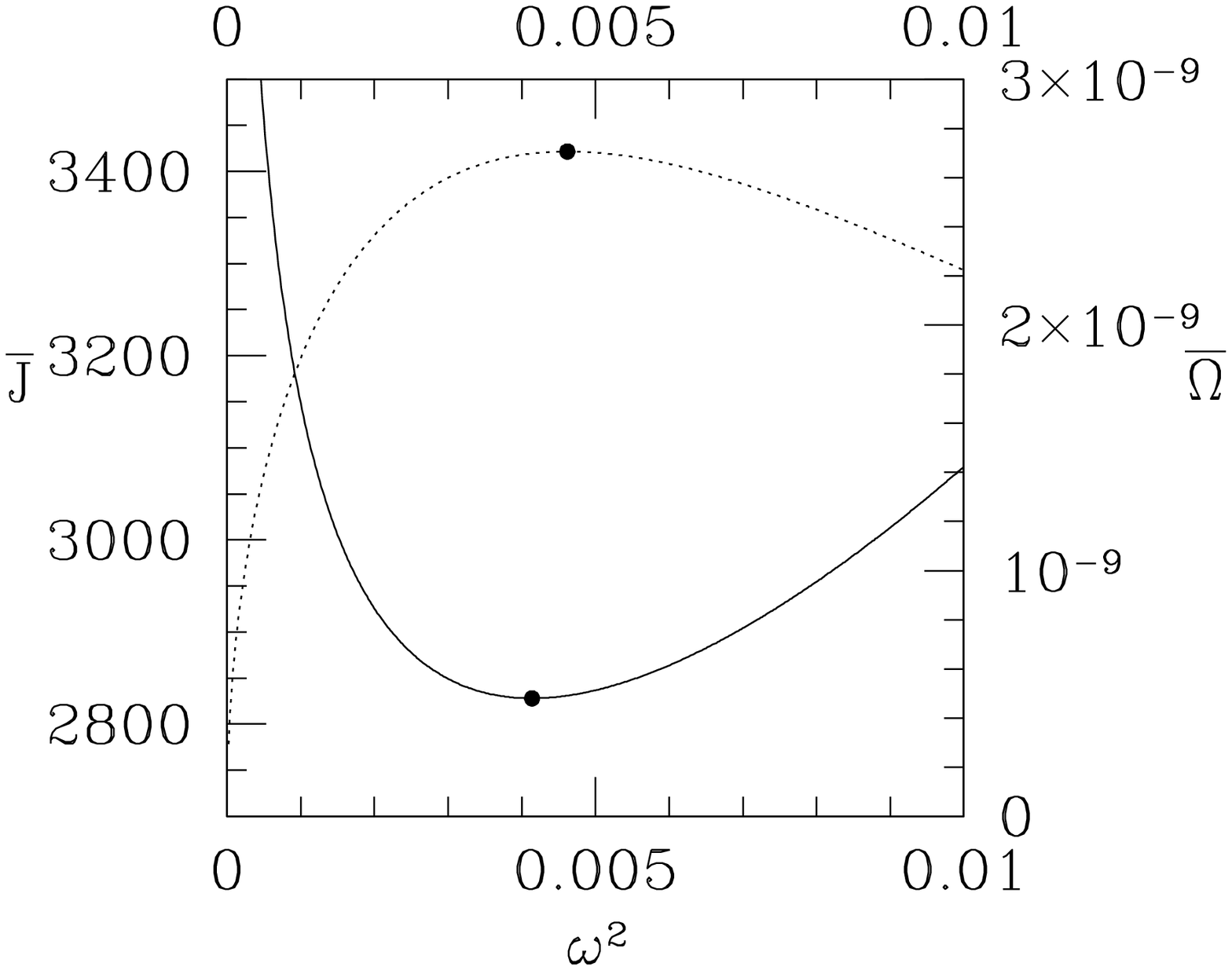}
\epsfxsize=3.5in
(d)~\epsffile{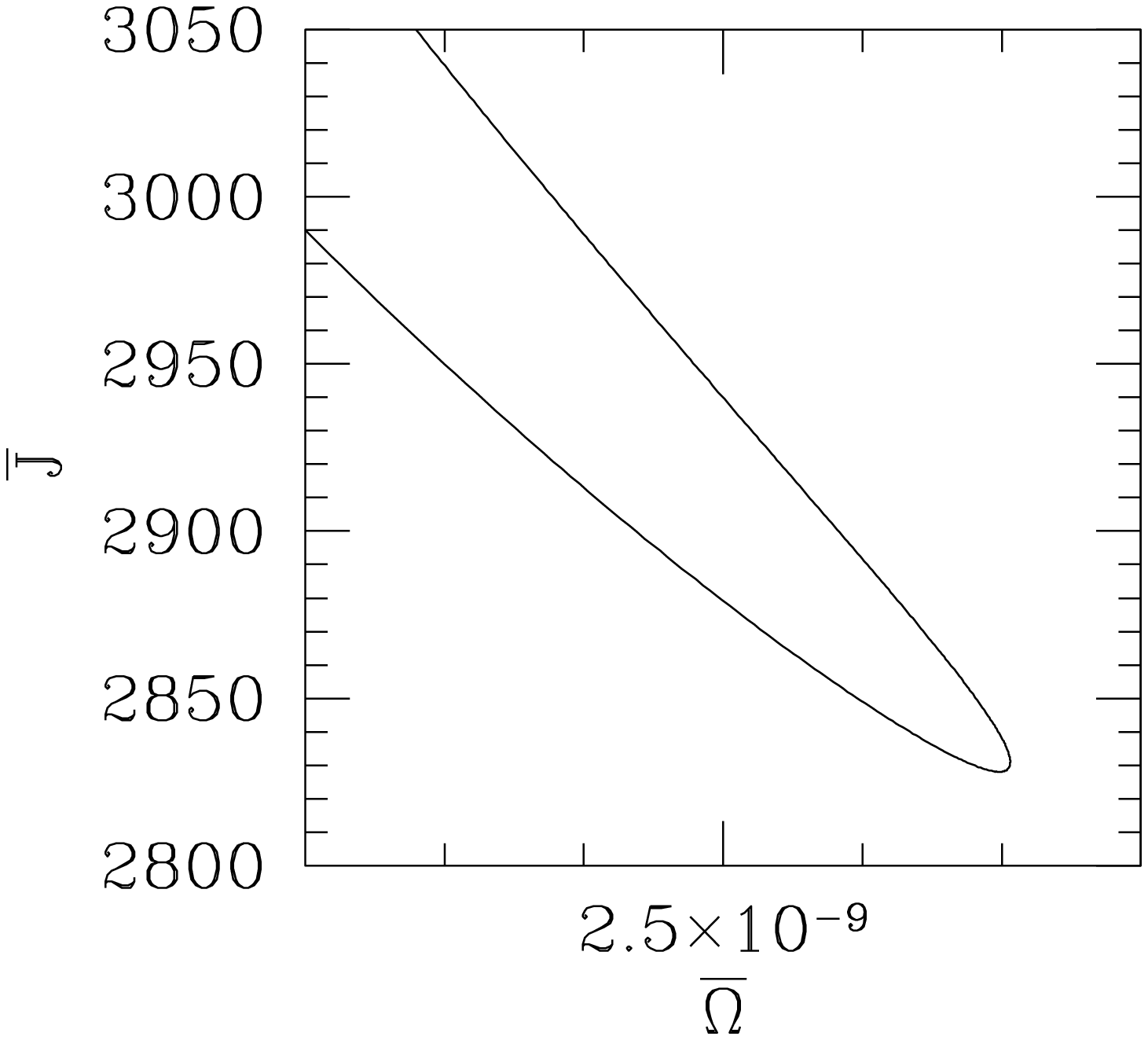}
\end{center}
\vspace*{-2mm}
\caption{(a) $\bar J$ (solid curve) and $\bar \Omega$ 
(dotted curve) as a function of 
$\omega^2$ along a fixed value of $\bar M~(=4.05)$ for $\epsilon=0.003$. 
Two solid circles indicate the minimum of $\bar J$ and maximum of 
$\bar \Omega$. (b) Relation between $\bar J$ and $\bar \Omega$ 
near the minimum of $\bar J$ for $\bar M=4.05$ and $\epsilon=0.003$. 
(c) The same as (a) but for $\bar M=3.90$. 
(d) The same as (b) but for $\bar M=3.90$.
}
\end{figure}

\begin{figure}[t]
\vspace*{0mm}
\begin{center}
\epsfxsize=3.5in
~\epsffile{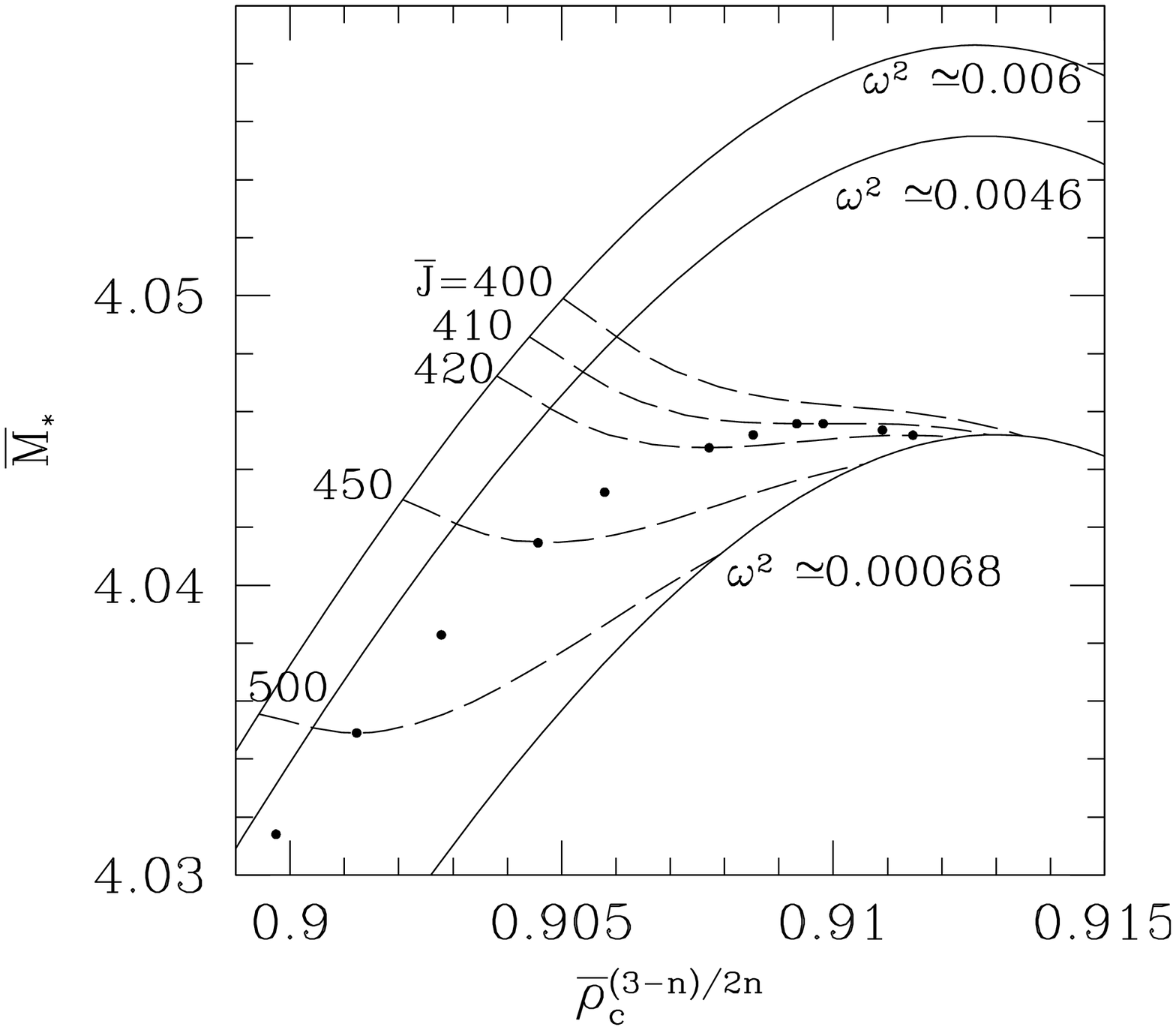}
\end{center}
\vspace*{-2mm}
\caption{Numerical mass-density relation for $\epsilon=0.003$. 
The dependence of $\bar M$ on $\bar \rho_{c}^{(3-n)/2n}$ for 
fixed values of $\omega^2$ is indicated by the solid lines. 
The dependence of $\bar M$ on $\bar \rho_{c}^{(3-n)/2n}$ for 
fixed values of $\bar J$ is shown by the dashed lines. 
The solid dots indicate the sequence of the turning points. 
}
\end{figure}

\begin{figure}[t]
\vspace*{0mm}
\begin{center}
\epsfxsize=3.5in
~\epsffile{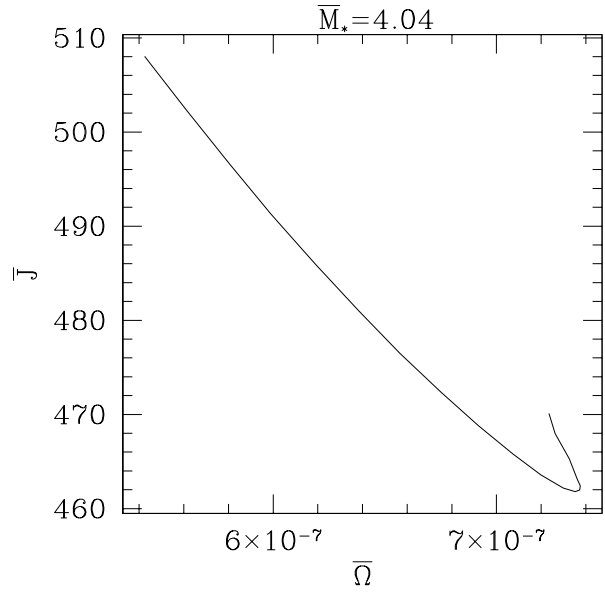}
\end{center}
\vspace*{-2mm}
\caption{Numerical relation between $\bar J$ (solid curve) and $\bar \Omega$ 
for a given value of $\bar M_*~(=4.04)$ for $\epsilon=0.003$. 
}
\end{figure}

\begin{figure}[t]
\vspace*{0mm}
\begin{center}
\epsfxsize=3.7in
~\epsffile{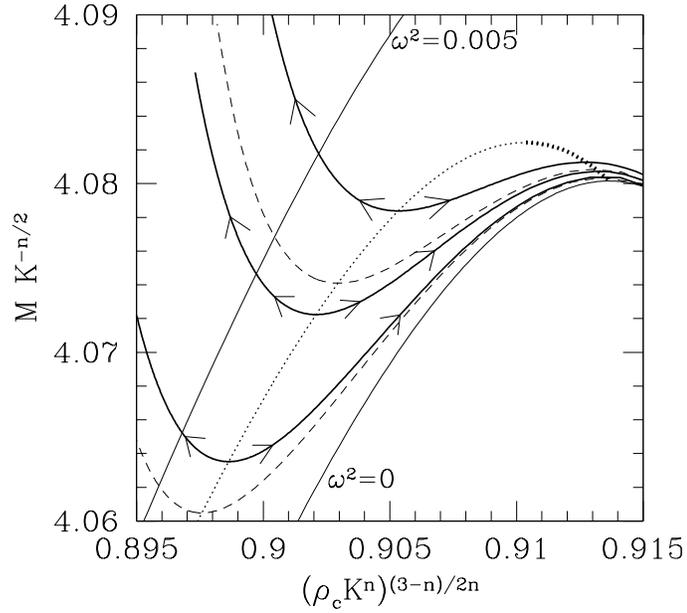}
\end{center}
\vspace*{-2mm}
\caption{Evolutionary tracks along curves of fixed 
$\bar J/\bar M^2(=28, 31$ and 35, thick solid lines from top to bottom) 
for $\epsilon=0.003$. The evolution must proceed toward increasing 
$\bar M$ as indicated by arrows. 
The thin solid lines denote fixed values of 
$\omega$ and the dashed lines show curves of fixed value of $\bar J(=500$ 
(upper) and 600 (lower)), used to determine the location of turning points 
(dotted line). The thin dotted line marks the threshold between 
increasing and decreasing orbital separation. 
The thick dotted line marks the onset of quasi-radial collapse.  
}
\end{figure}

\begin{figure}[t]
\vspace*{0mm}
\begin{center}
\epsfxsize=3.7in
~\epsffile{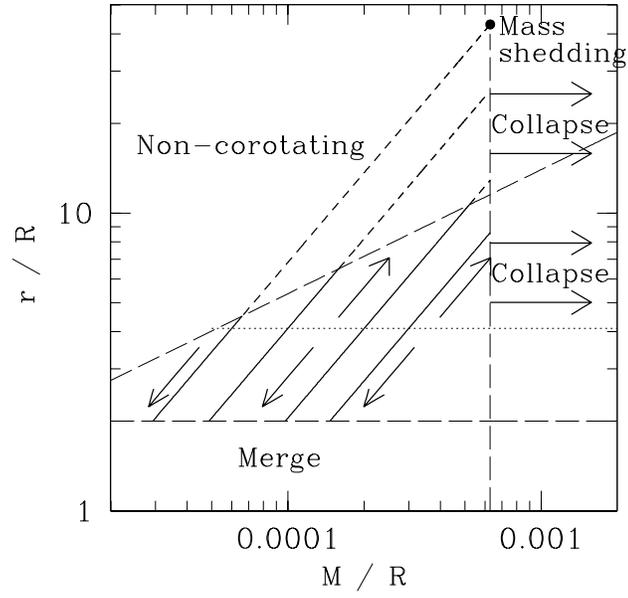}
\end{center}
\vspace*{-2mm}
\caption{Schematic diagram for the quasi-stationary evolution of 
supermassive binary stars of equal mass. 
The solid arrows indicate the direction of evolution as a result of 
cooling via thermal radiation. The solid and dashed lines indicate 
possible evolutionary paths. 
}
\end{figure}

\end{document}